\documentclass{rsproca_new}

\usepackage{lineno}
\usepackage{bm}
\usepackage{graphicx}
\usepackage{subcaption}
\usepackage{amsmath}
\usepackage{amssymb}
\usepackage[labelsep=period, labelfont=bf]{caption}
\usepackage{threeparttable}
\usepackage{hyperref}

\renewcommand{\figurename}{Fig.}
\newcommand{\maxnorm}[1]{\| #1 \|_\infty}
\newcommand{\fnorm}[1]{\| #1 \|_\text{F}}
\newcommand{\SRS}{\text{\tiny SRS}}
\newcommand{\SSI}{\text{\tiny SSI}}

\jname{rspa}
\Journal{Proc R Soc A\ }

\begin{document}

\title{Metric of Shock Severity}

\author{
Yinzhong Yan and Q.M. Li}

\address{Department of Mechanical, Aerospace and Civil Engineering, School of Engineering, The University of Manchester, Manchester M13 9PL, United Kingdom}

\subject{Acoustics, Applied Mathematics, Mechanical Engineering}


\corres{Q.M. Li\\
\email{qingming.li@manchester.ac.uk}}

\begin{abstract}
Shock response spectrum (SRS) is a widely accepted method for shock testing specification.
However, SRS, as a supremum, is over-conservative and cannot fully describe the relative severity for various shock environments.
This study introduces shock response matrix, from which shock severity infimum (SSI) can be extracted using singular value decomposition method.
Based on SRS and SSI, dual spectra are introduced to determine the range of shock severity between its supremum and infimum.
The evaluation of the relative severity of various shock signals is discussed and validated by finite element simulations.
\end{abstract}


\begin{fmtext}
\section{Introduction}

Mechanical shocks widely exist in many engineering applications, such as aerospace\cite{ECSS2015}, navy\cite{MIL-S-901D1989} and defence engineering\cite{810G}.
These mechanical shocks normally do not introduce damage to the main structures, but they may possibly cause major functional failures of on-board products, e.g., electronic and optical devices and other equipment, which would subsequently result in the partial or total loss of a mission.
Designed products should be tested accordingly in laboratories to ensure their reliability in mechanical shock environment during their service life.
However, a complete reproduction of field mechanical shock environment in a laboratory for product validation is almost impossible.
Unlike other simple impulsive or stationary vibration excitations, it is difficult to define the shock severity of a mechanical shock in time and frequency domains due to its complexity.
A metric is necessary for the measure of shock severity and the comparison between a real shock and the simulant shock reproduced in laboratory environments.
\end{fmtext}


\maketitle

The current metric, shock response spectrum (SRS), was proposed by Biot in 1932\cite{biot1932}.
SRS established a kind of measure for shock severity and shock equivalence in an elastic dynamics way.
It evaluates the modal responses using a standardized dynamical system (i.e. SDOF) and a selected maximum response quantity.
It has been shown that the maximum stress is proportional to the maximum particle velocity for a given material and model shape\cite{Hunt1960,Gaberson1969}.
For a given vibration mode, displacement, velocity and acceleration can be linked by the vibration frequency of the mode, and consequently, the stress can be correlated to the acceleration\cite{Gaberson2012}.
Therefore, the shock severity or its destructive potential is evaluated by the maximum structural response, which can be estimated by a given SRS using a response spectrum analysis method, e.g. absolute sum (ABSSUM) method\cite{Biot1933a, Biot1934a}. 
The estimated maximum stress can be directly compared to material strength\cite{ECSS2015} to define the damage level, so that the equivalence among various shocks can be established by their SRS curves.
The initial purpose of this method was to study the effects of earthquakes on buildings and provide a methodology for seismic resistant design.
Later this method was generalized for the analysis of various types of mechanical shocks.
SRS was first adopted as the standard testing specification in 1983\cite{MIL-STD-810D1983} to compare relative shock severity between measured and reproduced shocks.
To date, SRS is still the state-of-the-art in measuring shock severity in testing standards in various engineering sectors, such as aerospace (e.g. NASA-STD-7003A\cite{NASA-STD-7003A2011}), defence (e.g. MIL-STD-810G\cite{810G}) and navy (e.g. MIL-S-901D\cite{MIL-S-901D1989}).
However, the shock equivalence based on SRS is questionable.
If the concerned structure is not specified, the maximum response of the structure cannot be evaluated appropriately using only SRS.
More specifically, the SRS method only provides the supremum for the maximum response range of a given structure.
For example, two shock signals with very close SRS curves can lead to over 40\% difference of the maximum acceleration responses for a multiple degree of freedom (MDOF) system\cite{christian2014}.
Especially for complex devices and equipment with rich contents of middle and high natural frequencies, SRS may not accurately represent the shock severity.

Other attempts have been made to give a better metric of shock severity.
For example, pseudo-velocity shock response spectrum (PVSRS) was suggested to substitute the maximum absolute acceleration\cite{Gaberson2012}.
However, the PVSRS is proposed as a rule of thumb without strict analytical justification and has not been widely accepted.
The fatigue damage spectrum (FDS) intends to combine a fatigue model with the shock response spectrum\cite{VanBaren2015}.
Some 3D contour spectra were proposed to represent the dynamical characteristics for a packaging system by changing the traditional SDOF responding model to 2-DOF\cite{Yu2012a}, 3-DOF\cite{Xu2012a} and tilted support\cite{Chen2014c} models.
These response spectra could be useful in a specific area, but their general applications are limited and they are still based on the same concept of SRS.
Based on the authors' best knowledge, there are no other comparable mechanical methods that may offer a better measure of shock severity than the measure based on SRS.

This study introduces the shock response matrix and shock response contour, from which a new measure of shock severity, i.e. shock severity infimum (SSI), can be extracted using singular value decomposition (SVD) method.
SSI provides the infimum of shock severity for a given shock environment, which together with SRS, offers a dual-spectra bounding the range of shock severity.
The proposed method is validated and illustrated by analysing the response of a simple structure based on the finite element method (FEM).

\section{Shock Response Matrix and Contour}\label{section_src}

When a shock represented by an acceleration-time history excitation is applied to a series of SDOF models, the temporal acceleration response of each SDOF oscillator is calculated.
These responses constitute a shock response matrix, whose two coordinates are time and natural frequency of a series of SDOF oscillators.
It should be noted that the proposed method is also applicable when other response quantities are adopted, e.g., displacement and velocity.

\begin{figure}
	\centering
	\includegraphics[width=0.85\linewidth]{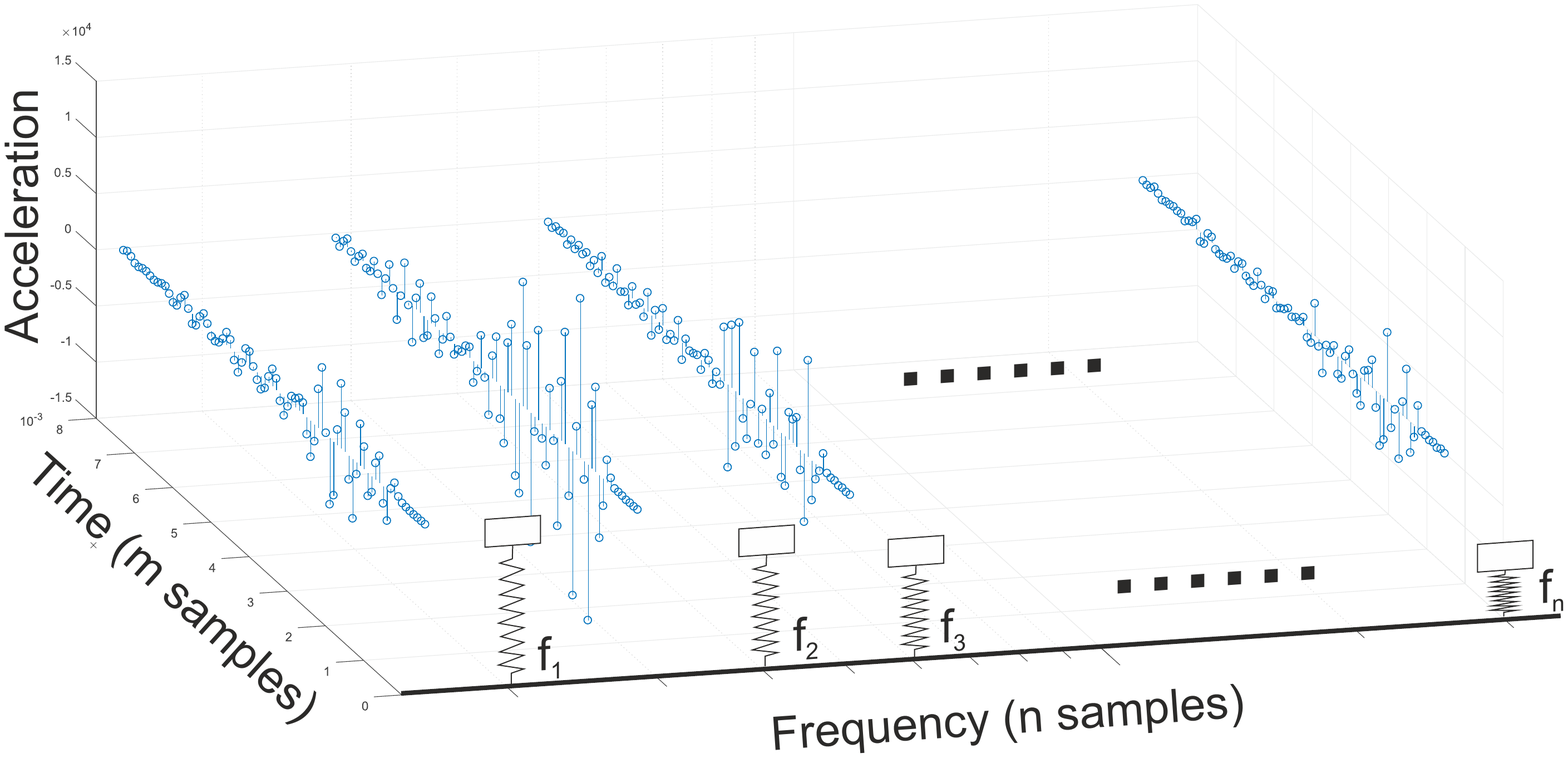}
	\caption{Diagram of shock response matrix $\bm{M}$}
	\label{diagram_src}
\end{figure}

A schematic graph of shock response matrix $\bm{M}$ is presented based on the SRS calculation algorithm.
As shown in Fig.\ref{diagram_src}, time series data ($m$ samples) of temporal responses of all $n$ SDOF oscillators are obtained, which are assembled into an $m\times n$ real matrix $\bm{M}$.
As required in standards, time series sample number is usually much larger than frequency series sample number, and therefore, $m\gg n$ in this case.
To be in line with the conventional SRS method, the element-wise magnitude of matrix $\bm{M}$ is introduced and represented by a new $m\times n$ matrix $\bm{N}$, i.e.,
\begin{equation}\label{avoid_small}
n_{ij}=|m_{ij}|.
\end{equation}

To show the acceleration responses of all considered SDOFs, the contour plot of the matrix $\bm{N}$ is drawn, which is named shock response contour (SRC) in this paper.
The frequencies and amplitudes are presented in logarithm scale in order to show signals' low-frequency character.
Large negative values may be produced during the use of logarithm expression if there are some very small components in matrix $\bm{N}$.
To focus only on severe responses, the data range shown in the SRC is consistent with its corresponding SRS curve, e.g., from 240 m/s$^2$ to 200,000 m/s$^2$ in this study.

When SRC is projected to the acceleration-frequency plane, the spectrum of the maximum response of the projected SRC is SRS, which is a vector consisting of the maximum of each time series response in matrix $\bm{N}$, i.e.,
\begin{equation}\label{srs_definition}
\bm{v}_{\SRS}=[\max_{1\leq i\leq m} n_{ij}]^\top
\end{equation}
The main problem of SRS is the missing of phase information of peak acceleration responses, which can be visualized by the bright line in SRC.

\begin{figure}
	\centering
	\includegraphics[width=0.75\linewidth]{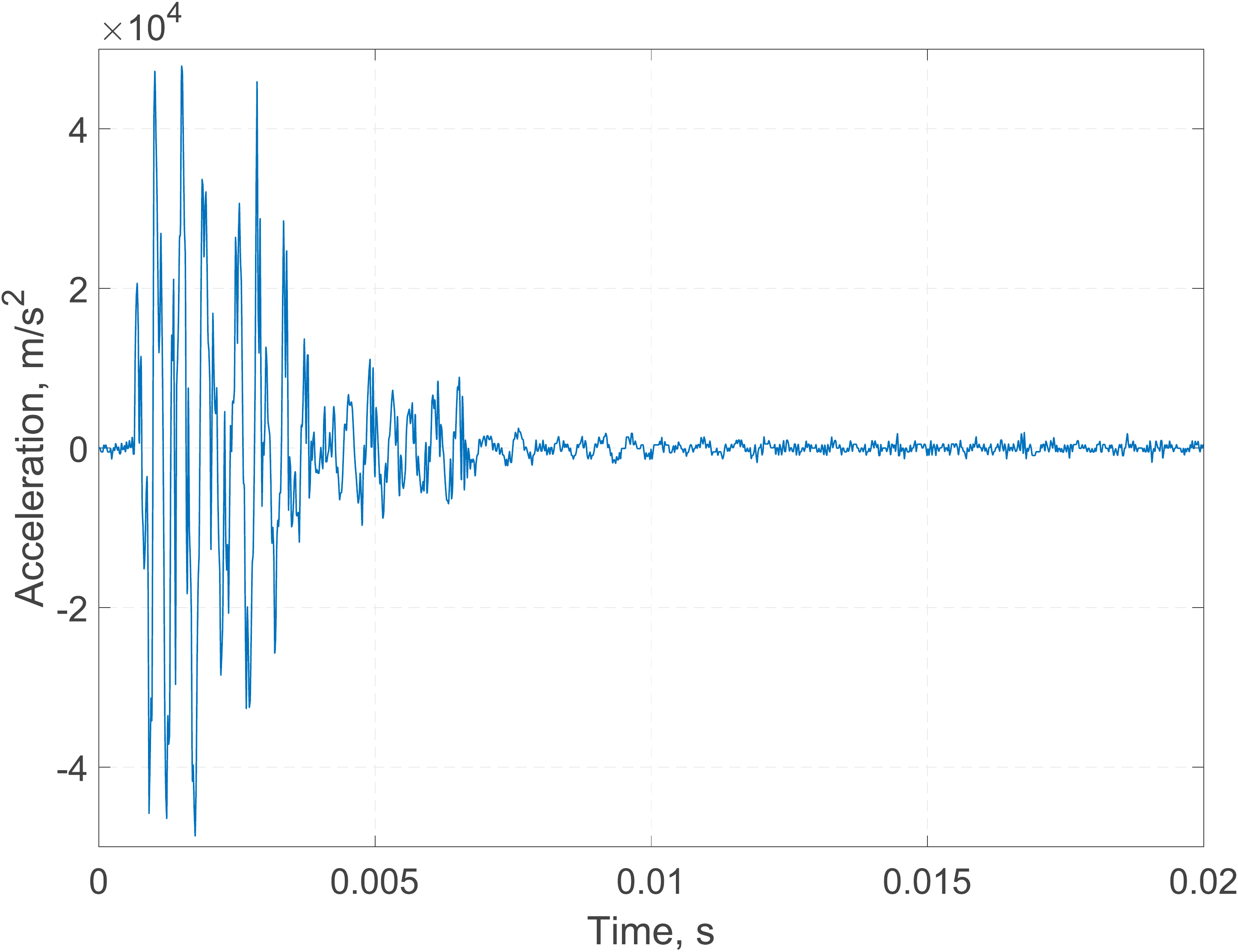}
	\caption{Acceleration-time history of `RVS'}
	\label{rvs}
\end{figure}

\begin{figure}
	\centering
	\begin{subfigure}[b]{0.49\linewidth}
		\includegraphics[width=\linewidth]{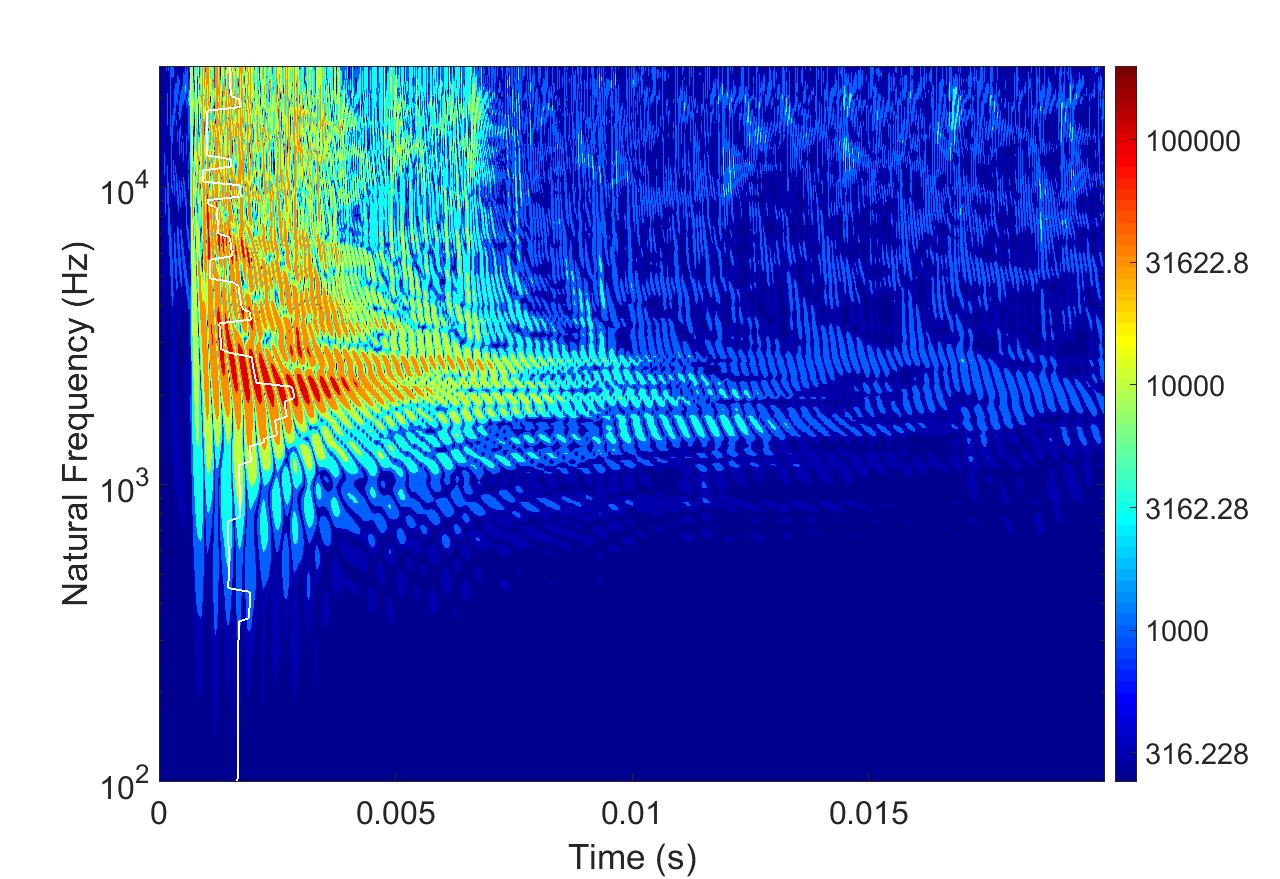}
		\caption{SRC of shock `RVS'}
		\label{rvs_src}
	\end{subfigure}
	\begin{subfigure}[b]{0.49\linewidth}
		\includegraphics[width=\linewidth]{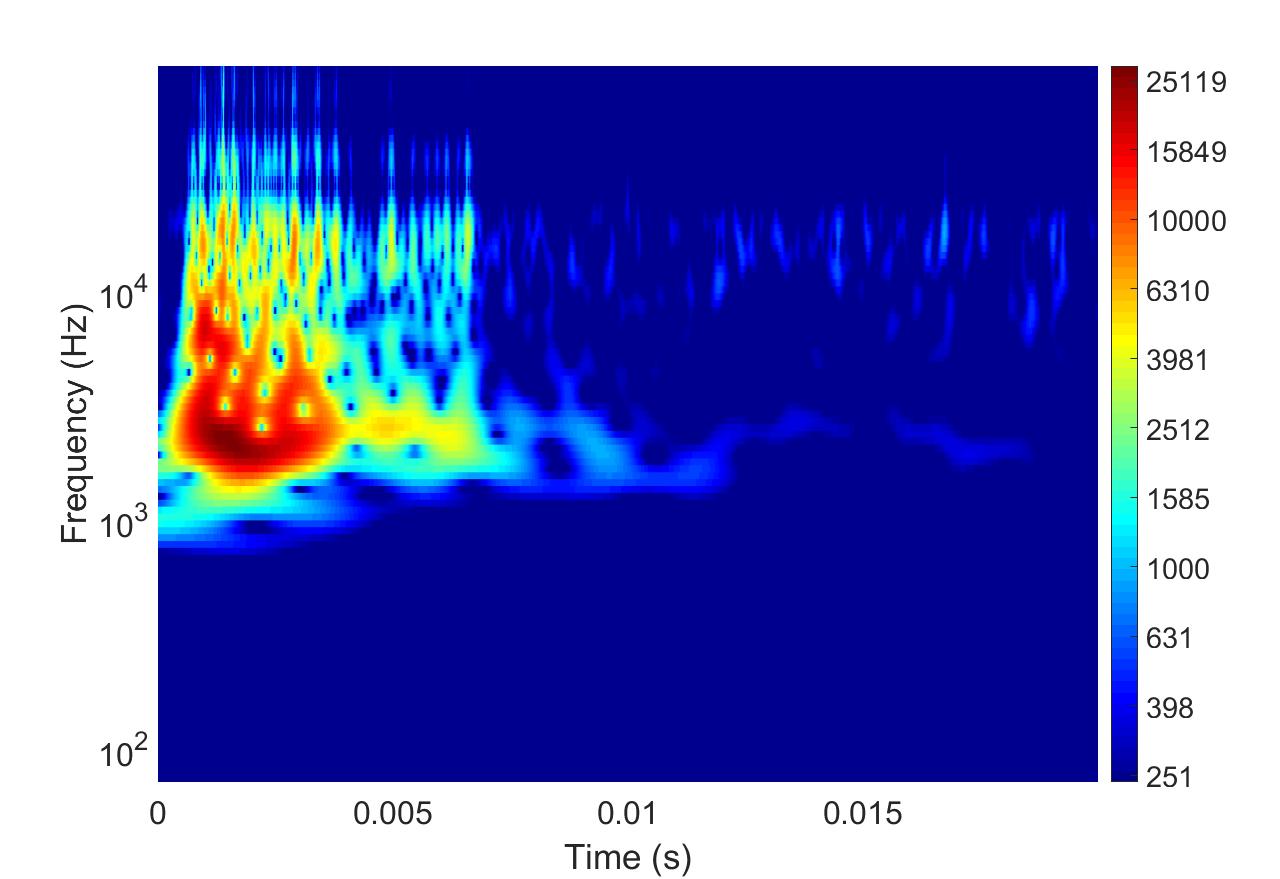}
		\caption{CWT of shock `RVS'}
		\label{rvs_cwt}
	\end{subfigure}
	\caption{Comparison between SRC and CWT for `RVS' shock signal}
	\label{rvs_src_cwt}
\end{figure}

Fig.\ref{rvs} shows the acceleration-time history of shock `RVS', which is a near-field pyroshock measured in the separation stage of an unnamed re-entry rocket vehicle\cite{irvine2010}.
The SRC of `RVS' illustrated in Fig.\ref{rvs_src} can provide rich information for its temporal structure, and hence, could be potentially used as a time-frequency analysis method with considering the response of each SDOF oscillator to the given shock.
SRC can be compared to the frequently-used method for the analysis of a non-stationary signal in the time-frequency domain, continuous wavelet transform (CWT), which projects a signal on a series of zero-mean basis functions derived from an elementary function by dilations and translations.
The CWT scalogram of `RVS' with analytical Morlet wavelet is plotted in Fig.\ref{rvs_cwt} using Matlab, against the corresponding SRC in Fig.\ref{rvs_src}.
The pattern of colour scale is clearer in SRC since it reveals the features of the mechanical system, e.g. resonant response.
The relationship between SRC and CWT is comparable to the relationship between SRS and fast Fourier transform (FFT) in terms of their mechanical and mathematical representations.
Further studies with focuses on the application of SRC are necessary but are outside the scope of this paper.

\section{Feature Extraction Using Singular Value Decomposition}

The equivalence of shock severity can be measured by a better method based on the features of matrix $\bm{N}$.
SRS is the absolute maximum feature of matrix $\bm{N}$, but it totally ignores the phase information.
Beside absolute maximum, there are other available matrix analysis methods that can be adopted to extract frequency features from matrix $\bm{N}$.
In this paper, the singular value decomposition (SVD) will be applied.

\subsection{Singular Value Decomposition}

\begin{figure}[t!]
	\centering
	\includegraphics[width=0.75\linewidth]{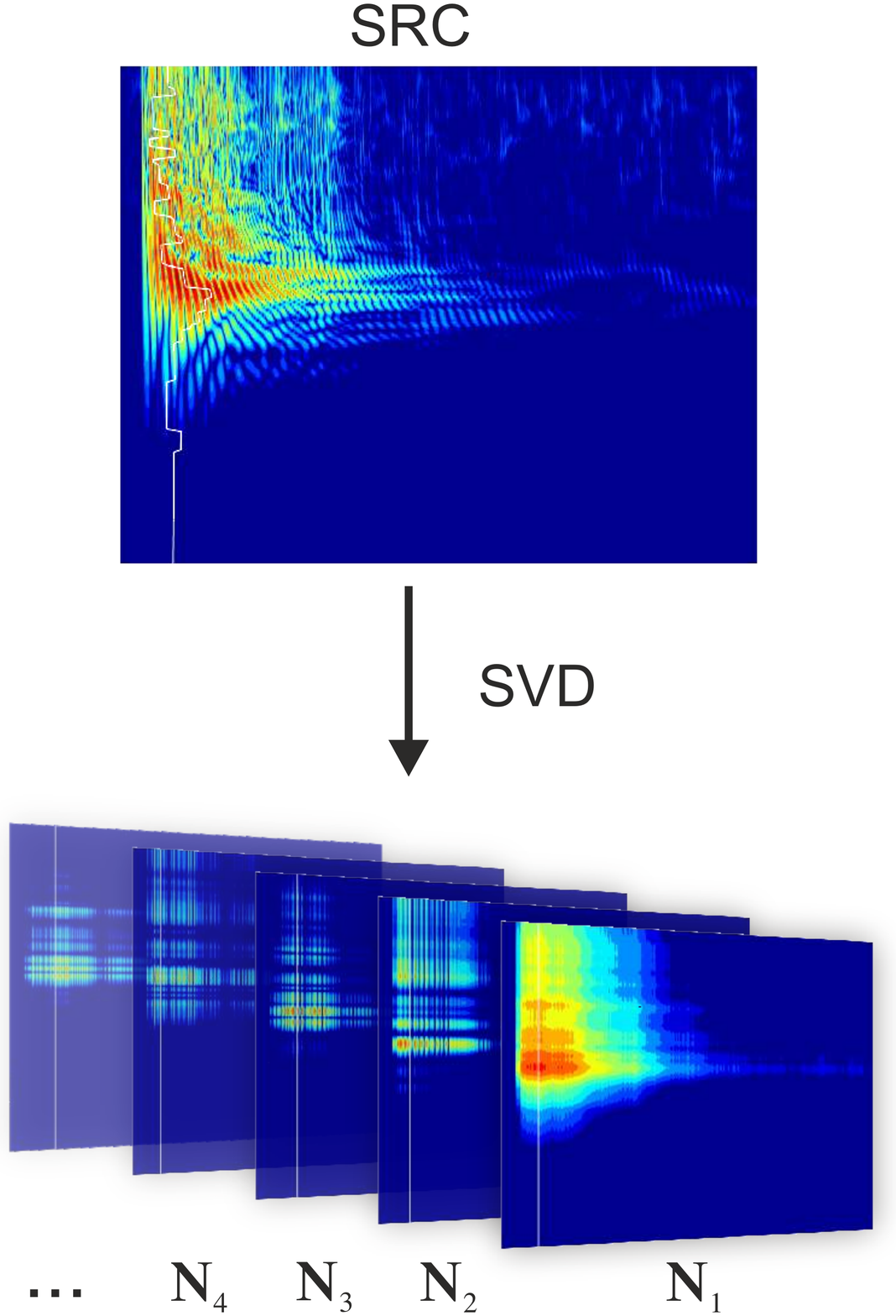}
	\caption{Diagram of SRC decomposition}
	\label{diagram_svd}
\end{figure}

The SVD method\cite{Strang2016} can be used to decompose a rank $r$ matrix $\bm{N}$ into $r$ rank-one matrices $\bm{N}_k =[n_{ij}^k]_{m\times n} \ (k=1,2,\cdots r)$, which can be sorted by their singular values in a descending order, i.e.,
\begin{equation}\label{svd_decomposition}
\bm{N}=\bm{U}\bm{\Sigma}\bm{V}^\top=\sum_{k=1}^r \sigma_k \bm{u}_k \bm{v}_k^{\top}=\sum_{k=1}^r \bm{N}_k
\end{equation}
where $\bm{U}$ is an $m\times m$ orthogonal matrix, $\bm{\Sigma}$ is a diagonal $m \times n$ matrix with non-negative real numbers on the diagonal, and $\bm{V}$ is an $n\times n$ orthogonal matrix.
The $\bm{u}_k$ and $\bm{v}_k$ are the $k$th column of $\bm{U}$ and $\bm{V}$, corresponding to the $k$th singular vector in time and frequency domains, respectively, and $\sigma_k$ is the $k$th-order singular value.
The SRC decomposition and the contour plot of each decomposed component ($\bm{N}_k$) are shown in Fig.\ref{diagram_svd}.

\subsection{Shock Severity Infimum}\label{Separability}

The matrix $\bm{N}$ can be generally considered as the discrete equivalence of a bivariate function $g(t,f)$ for the shock responses of SDOF oscillators, at regularly spaced discrete values of time $t$ and natural frequency $f$.
If the shock response bivariate function $g(t,f)$ can be factorized into two single-variate functions,
\begin{equation}\label{separable_function}
g(t,f)=u(t) \cdot v(f)
\end{equation}
The separated function in frequency domain $v(f)$ would provide a better description for the metric of shock severity because phase difference will not be an issue here.

However, the bivariate function corresponding to matrix $\bm{N}$ is normally not a separable function.
Therefore, this study intends to find a rank-one (separable) matrix $\widehat{\bm{N}}$ with the maximum fit to $\bm{N}$.
The mathematical expression of this minimization problem is given by
\begin{equation}\label{matrix_approximation_problem}
\text{minimize} \quad \text{over} \  \widehat{\bm{N}} \quad  \fnorm{\bm{N}-\widehat{\bm{N}}}  \quad \text{subject to} \quad \text{rank}(\widehat{\bm{N}})=1
\end{equation}
where $\fnorm{\cdot}$ is the Frobenius norm.
According to the Eckart–Young–Mirsky theorem\cite{Eckart1936}, $\widehat{\bm{N}}=\bm{N}_1$, where $\bm{N}_1$ is the first SVD component of $\bm{N}$.
Thus, $\bm{N}_1$ can be associated with the discretized separable function $g(t,f)$ in Eq.(\ref{separable_function}) in a matrix expression of $[g(t,f)]$.

To be in line with the definition of SRS, a new concept of shock severity infimum (SSI) of matrix $\bm{N}$ is introduced by projecting matrix $\bm{N}_1$ to the acceleration-frequency plane, i.e.,
\begin{equation}\label{n1}
\bm{v}_\SSI=[\max_{1\leq i\leq m} n^1_{ij}]^\top=\sigma_1 \maxnorm{\bm{u}_1} \bm{v}_1
\end{equation}
where $\maxnorm{\cdot}$ is the maximum norm.
The $\bm{u}_\SSI$ is defined correspondingly as $\bm{u}_1$ normalized by its maximum value, i.e.,
\begin{equation}
\bm{u}_\SSI= \dfrac{\bm{u}_1}{\|\bm{u}_1\|_\infty},
\end{equation}
so that $\bm{N}_1$ can be re-expressed as
\begin{equation}\label{separable_function_discrete}
\bm{N}_1=\bm{u}_{\SSI} \cdot \bm{v}_\SSI^\top=[u(t)] \cdot [v(f)]^\top=[g(t,f)].
\end{equation}
It should be noted that the SSI of matrix $\bm{N}$ is the SRS of matrix $\bm{N}_1$.

SSI is not a real description, but an approximation of shock severity metric, as SSI is extracted from $\bm{N}_1$ rather than $\bm{N}$.
Error of approximation comes from the difference between $\bm{N}_1$ and $\bm{N}$.
The relative residual error $\alpha$ of this approximation problem can be evaluated by
\begin{equation}\label{relative_residual_error}
\begin{split}
\alpha &= \bigg( \dfrac{\fnorm{\bm{N}-\bm{N}_1}}{\fnorm{\bm{N}}} \bigg) ^2\\
&= \bigg( \dfrac{ \fnorm{\sum_{k=2}^r \bm{N}_k}}{\fnorm{\bm{N}}} \bigg) ^2\\
&= \dfrac{\sum_{k=2}^{r} \sigma_k^2}{\sum_{k=1}^{r} \sigma_k^2}\\
&= 1-\dfrac{\sigma_1^2}{\sum_{k=1}^{r} \sigma_k^2}
\end{split}
\end{equation}
according to the property of SVD method and Forbenius norm.
$\alpha$ is a measure of the ``distance" of the system $\bm{N}$ from a separability function.
For example, $\alpha$ has been used as the degree of inseparability of a bivariate function in the study of the dynamic ripples in ferret primary auditory cortex\cite{Depireux2001}.
$\alpha$ value close to zero indicates good representation with just one decomposed component, i.e. just SSI can well estimate the shock severity.
In this case, SSI could possibly replace the current SRS technique.
$\alpha$ value approaching unity indicates an increase of inseparability, where only SSI can hardly estimate the shock severity.
For the latter case, both SRS and SSI should also be considered together as a dual spectra technique to bound the real shock severity.

\section{Metric of Shock Severity}

\subsection{Structural response based on modal analysis}

For a linear $n$-degree of freedom system, its $n\times n$ modal matrix of mass normalized eigenvector $\bm{\Phi}=[\phi_{ij}]$ can be calculated, where $\phi_{ij}$ is the modal displacement\cite[ch.11]{Clough2003a}.
Based on the modal analysis method, the actual structural acceleration response at coordinate $i$, i.e., $\bm{a}_i$, is considered as the superposed acceleration-time history response of a series of vibration modes\cite{alexander2009},
\begin{equation}
\bm{a}_i=\bm{M}\bm{x}_i
\end{equation}
where $\bm{x}_i=\bm{p}\circ \bm{\phi}_i$ is the Hadamard product of modal participation factor vector $\bm{p}$ and the eigenvector coefficient $\bm{\phi}_i$ at coordinate $i$.
The vector $\bm{\phi_i}$ is taken from the $i$th row of matrix $\bm{\Phi}$, and usually indicate the concerned location on structure.

The maximum absolute acceleration response has been used to calculate the maximum stress in a structure\cite[p.~477]{ECSS2015}, thus, the maximum acceleration response $\maxnorm{\bm{a}_i}$ can be used to represent the effect of shock severity.
It is evident that the maximum acceleration response calculated from matrix $\bm{N}$ is slightly higher than the maximum acceleration response calculated from matrix $\bm{M}$, according to their respective definitions,
\begin{equation}\label{inequality1}
\maxnorm{\bm{a}_i}=\maxnorm{\bm{M} \bm{x}_i} \lessapprox \maxnorm{\bm{N} \bm{x}}
\end{equation}
where $\bm{x}$ is the entrywise absolute values of vector $\bm{x}_i$.
However, inequality relation based on $\bm{M}$ is difficult for analysis.
Therefore, its corresponding absolute matrix $\bm{N}$ is discussed instead.

The time history response $\bm{N} \bm{x}$ is a linear combination of weighted singular vector $\bm{u}_k$ in time domain, i.e.,
\begin{equation}\label{time_combination}
\begin{split}
\bm{N} \bm{x} & =\bm{N}_1 \bm{x} + \sum_{k=2}^{r} \bm{N}_k \bm{x}\\
& =\sigma_1 \bm{u}_1 \bm{v}_1 ^\top \bm{x} + \sum_{k=2}^{r} \sigma_k \bm{u}_k \bm{v}_k ^\top \bm{x} \\
& = (\sigma_1 \bm{v}_1 ^\top \bm{x}) \bm{u}_1 + \sum_{k=2}^{r} (\sigma_k \bm{v}_k ^\top \bm{x}) \bm{u}_k \\
\end{split}
\end{equation}
$\bm{u}_1$ is the most important singular vector due to its dominant weight and usually the only positive singular vector of $\bm{N}$ in time domain.
By applying Eckart–Young–Mirsky theorem\cite{Strang2016} as shown in Appendix \ref{proof_trend}, the temporal vector $(\sigma_1 \bm{v}_1 ^\top \bm{x}) \bm{u}_1$ approximates the $\bm{N} \bm{x}$ in a least square sense, i.e.,
\begin{equation}\label{trend}
\| \bm{N}\bm{x}-(\sigma_1 \bm{v}_1 ^\top \bm{x}) \bm{u}_1 \|_2 \leq \sigma_2 \| \bm{x} \|_2.
\end{equation}
$(\sigma_1 \bm{v}_1 ^\top \bm{x}) \bm{u}_1$ usually depicts the average trend of the time history response of $\bm{N}\bm{x}$, while the rest of weighted singular vectors, $(\sigma_k \bm{v}_k ^\top \bm{x}) \bm{u}_k$ in time domain, contribute to local oscillations around $(\sigma_1 \bm{v}_1 ^\top \bm{x}) \bm{u}_1$.
The maximum of $\bm{N} \bm{x}$ is always larger than the maximum of $\bm{N}_1 \bm{x}$ according to calculation observations.
Therefore,
\begin{equation}\label{inequality2}
\bm{v}_\SSI^\top\bm{x}=\maxnorm{\bm{N}_1 \bm{x}} \leq \maxnorm{\bm{N} \bm{x}} \leq \bm{v}_\SRS^\top \bm{x}
\end{equation}
according to the proofs in Appendix \ref{proof_inequality2}.
The method to calculate the supremum of maximum response with SRS ($\bm{v}_\SRS$) is termed as absolute response spectrum analysis method\cite{ECSS2015,alexander2009}, which is the most conservative prediction according to Eq.(\ref{inequality2}).

The left inequality relationships in (\ref{inequality2}) is the main assumption based on numerous numerical calculation observations, which will be demonstrated by examples in Section \ref{case}.

\subsection{Margin between SSI and SRS}

The margin between infimum and supremum of a shock can be calculated by the ratio vector $\bm{l}$ in dB, i.e. calculating the amplitude ratio between SRS and SSI in logarithm scale,
\begin{equation}\label{margin_db}
l[j]= 20 \lg (\dfrac{v_\SRS[j]}{v_\SSI[j]}) \ \text{dB}
\end{equation}
where $j$ represents the $j$th discrete frequency.
This frequency-dependent margin can show the uncertainty of SRS-based shock severity over the entire frequency range.
A smaller margin $\bm{l}$ indicates better representativeness of shock severity by either SRS or SSI.
Instead, a larger margin $\bm{l}$ indicates the uncertainty of the SRS to represent shock severity, for which other complementary methods, e.g., SSI, is necessary for the description of the shock severity.

\subsection{Uncertainty in Shock Testing}

\begin{figure}
	\centering
	\includegraphics[width=0.75\linewidth]{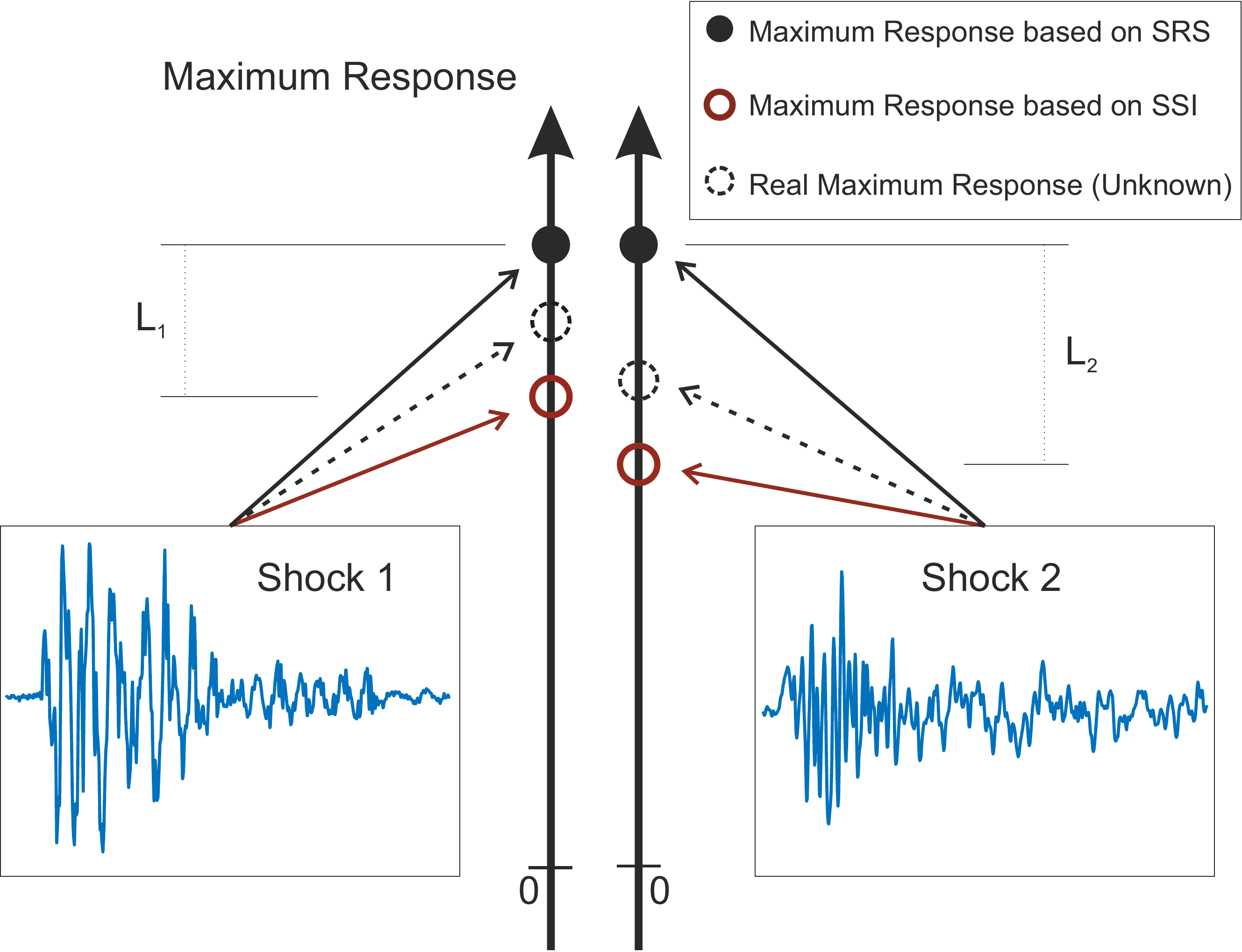}
	\caption{Measure of shock severity using SRS and SSI methods}
	\label{measure_severity}
\end{figure}

SRS is commonly used to evaluate the relative severity of shocks produced in field and laboratory environments in various industrial standards\cite{ECSS2015,810G,christian2014}.
Compared to the real maximum response, Fig.\ref{measure_severity} shows the measure of shock severity via SSI and SRS methods.
The real maximum response under two given shock excitations can not be calculated without the modal information of a specific structure.
Even if the SRS curves of two shocks are equivalent, it only indicates their severity equivalence in terms of the equivalence between their supremum responses.
The relationships between the magnitudes of their maximum responses to the two shocks are still unknown, which causes the uncertainties for the establishment of the equivalence between two shocks based on SRS.

\begin{figure}
	\centering
	\includegraphics[width=\linewidth]{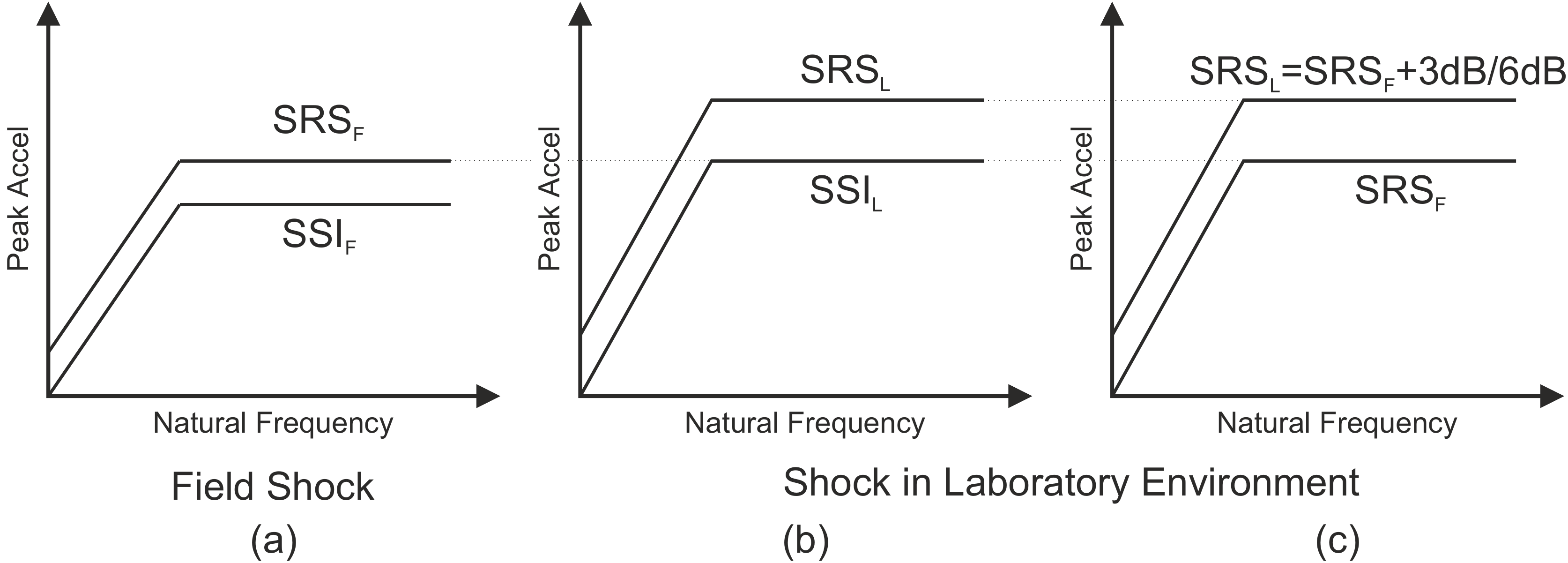}
	\caption{Fully conservative testing in a laboratory environment, (a) field shock, (b) laboratory shock with margin required by dual spectrum; (c) laboratory shock with 3dB/6dB margin. The subscript `F' and `L' represent the field shock and the laboratory shock, respectively.}
	\label{fully_conservative}
\end{figure}

The infimum and supremum of the maximum response to a shock can be represented by its SSI and SRS, respectively.
The range of all possible maximum responses can be perceived by the dual spectrum in an explicit way.
However, even the dual spectrum of both the field and laboratory testing shocks are the same, the response of equipment under field shock may still differ from the laboratory testing response, which may lead to over- or under-testing.
A fully conservative way to conduct laboratory shock testing is to make the SSI$_\text{L}$ of the laboratory testing shock to be equal to or higher than the SRS$_\text{F}$ of the field shock, as shown in Fig.\ref{fully_conservative}(b), i.e., the infimum of maximum response of the laboratory testing shock is no less than the supremum of maximum response of the field shock.
If equipment can pass such a test in a laboratory environment, then it shall also survive under the field shock.

As a comparison, a traditional method suggested by US Army 810G standard\cite{810G} is to apply a margin over the SRS$_\text{F}$ of field shocks, as shown in Fig.\ref{fully_conservative}(c), which considers the stochastic variability in the environment and the uncertainty in any predictive methods employed.
The margin is mainly suggested by engineering experience and statistical rationale, e.g., P95/50 rule\cite{Yunis2007}.
For mechanical shock, a 3dB or 6dB margin is normally added, depending on the degree of test level conservativeness desired\cite[p.~383]{810G}, while for pyroshock this margin is often set to 6dB.
A comparison between $\bm{l}$ margin and ordinary margin in a case study will be discussed in Section \ref{case}.

\section{Case Study}\label{case}

\subsection{Shock Signal Preparation and Analysis}

\begin{figure}[t]
	\centering
	\begin{subfigure}[b]{0.49\linewidth}
		\includegraphics[width=\linewidth]{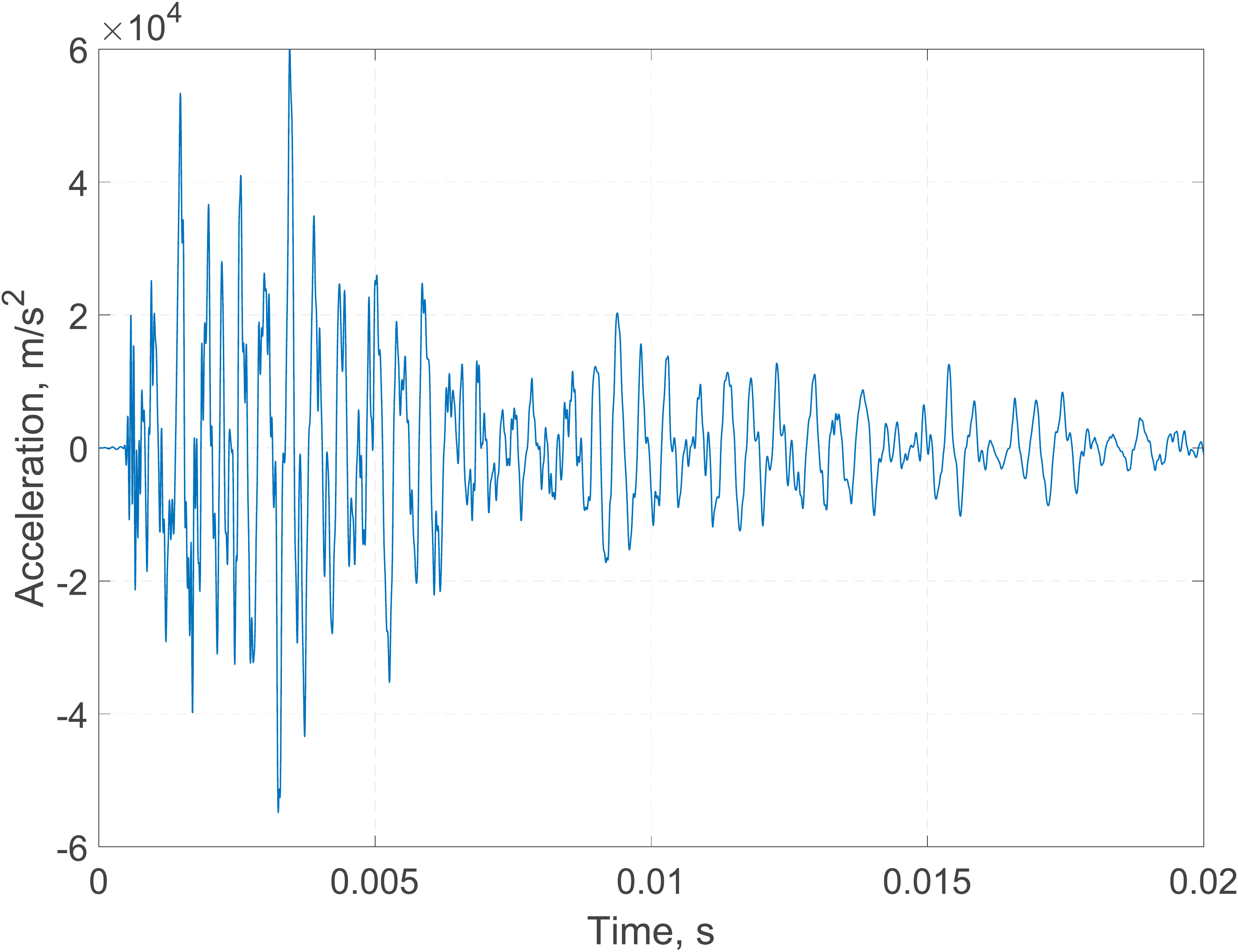}
		\caption{`MIS'}
		\label{mis}
	\end{subfigure}
	\begin{subfigure}[b]{0.49\linewidth}
		\includegraphics[width=\linewidth]{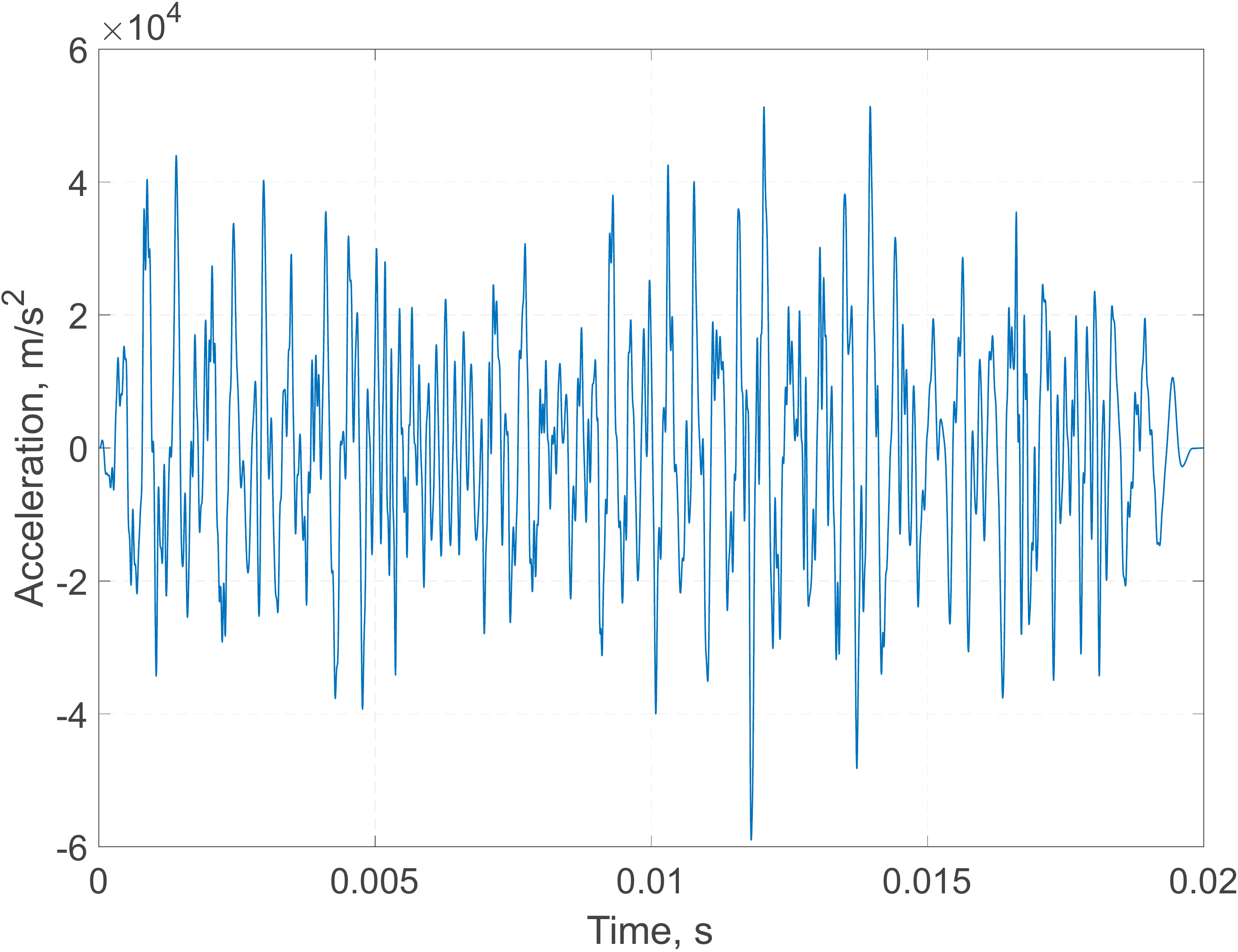}
		\caption{`RAS'}
		\label{ras}
	\end{subfigure}
	\begin{subfigure}[b]{0.49\linewidth}
		\includegraphics[width=\linewidth]{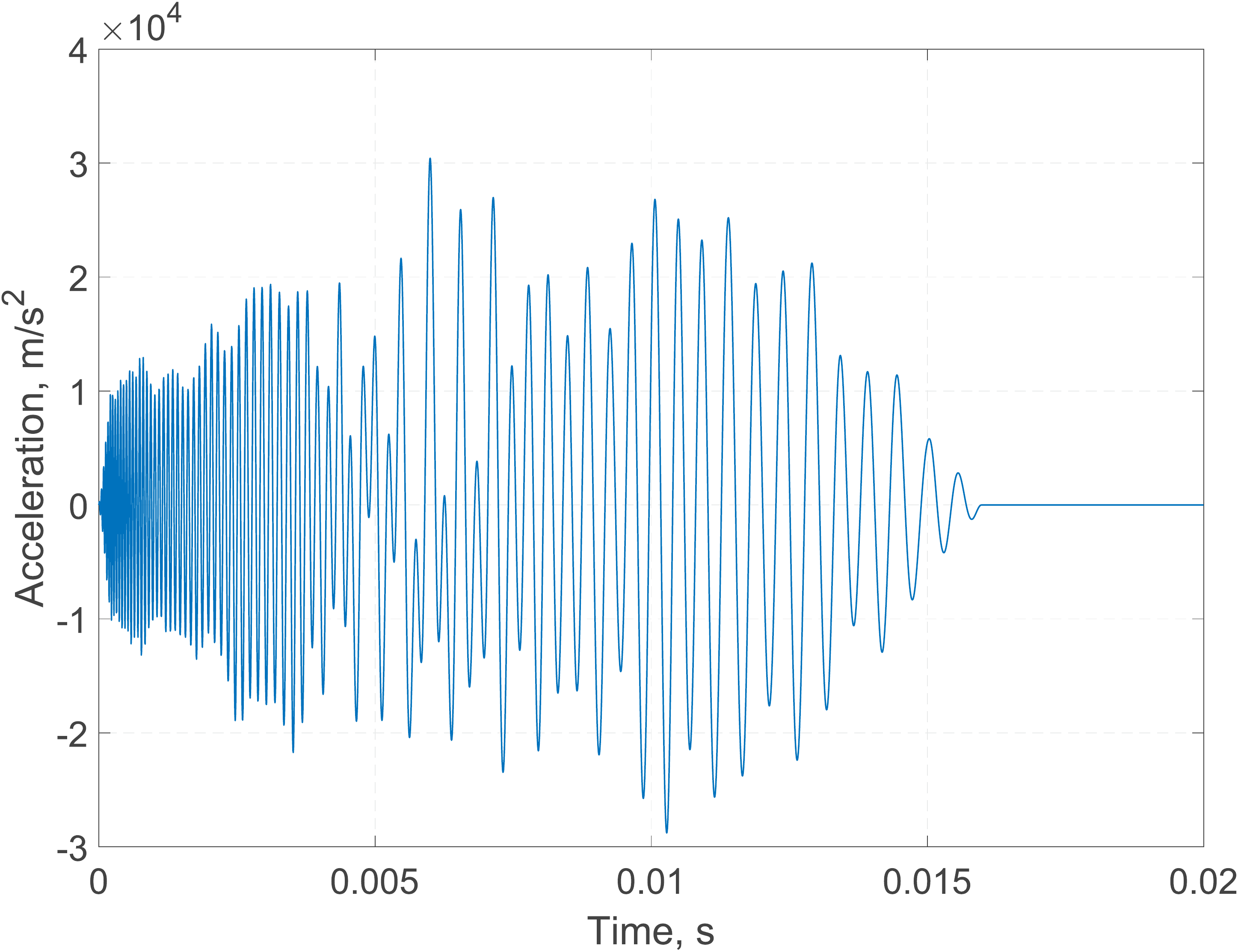}
		\caption{`RSS'}
		\label{rss}
	\end{subfigure}
	\caption{Time history of three shock signals `MIS', `RAS' and `RSS'}
	\label{time_history_compare}
\end{figure}

\begin{figure}[htpb!]
	\centering
	\includegraphics[width=0.75\linewidth]{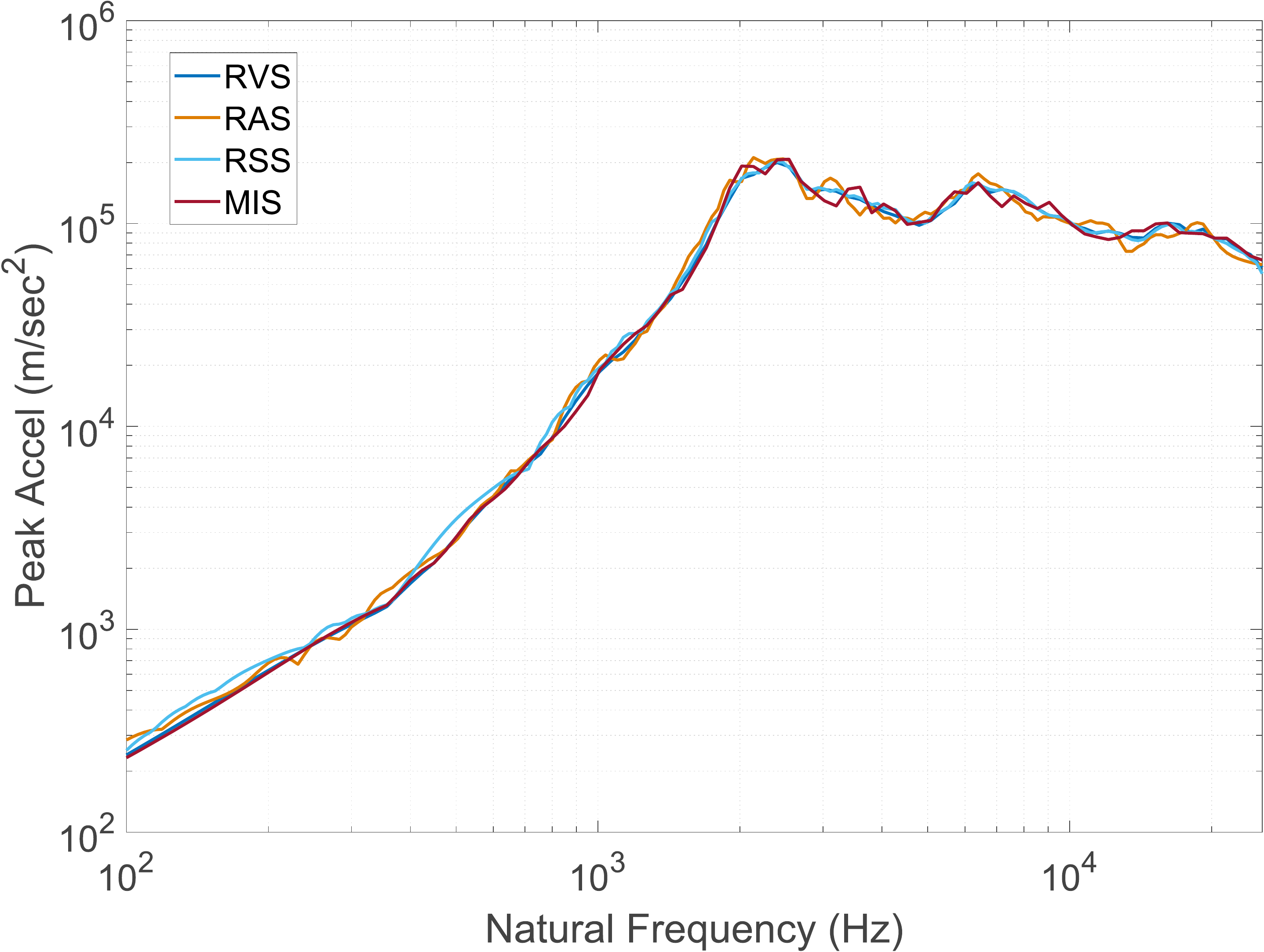}
	\caption{Comparison of SRS curves of four shock signals (`RVS', `RAS', `RSS' and `MIS')}
	\label{srs_compare}
\end{figure}

Three more shocks are synthesised through different mechanisms but having almost the same SRS of the pyroshock `RVS' described in Section \ref{section_src}.
The mechanical impact shock `MIS' is synthesised with the net zero displacement filter method using a field shock generated by metal-metal impact in laboratory environment\cite{Yan2019}.
The random arrangement shock `RAS' and the reverse sine sweep shock `RSS' are synthesised with Tom Irvine's Matlab script\cite{irvine2018}.
The time histories and SRS curves (Q=10) of these shock signals are shown in Figs.\ref{time_history_compare} and \ref{srs_compare}, respectively.

\begin{figure}
	\centering
	\begin{subfigure}{0.48\linewidth}
		\includegraphics[width=\linewidth]{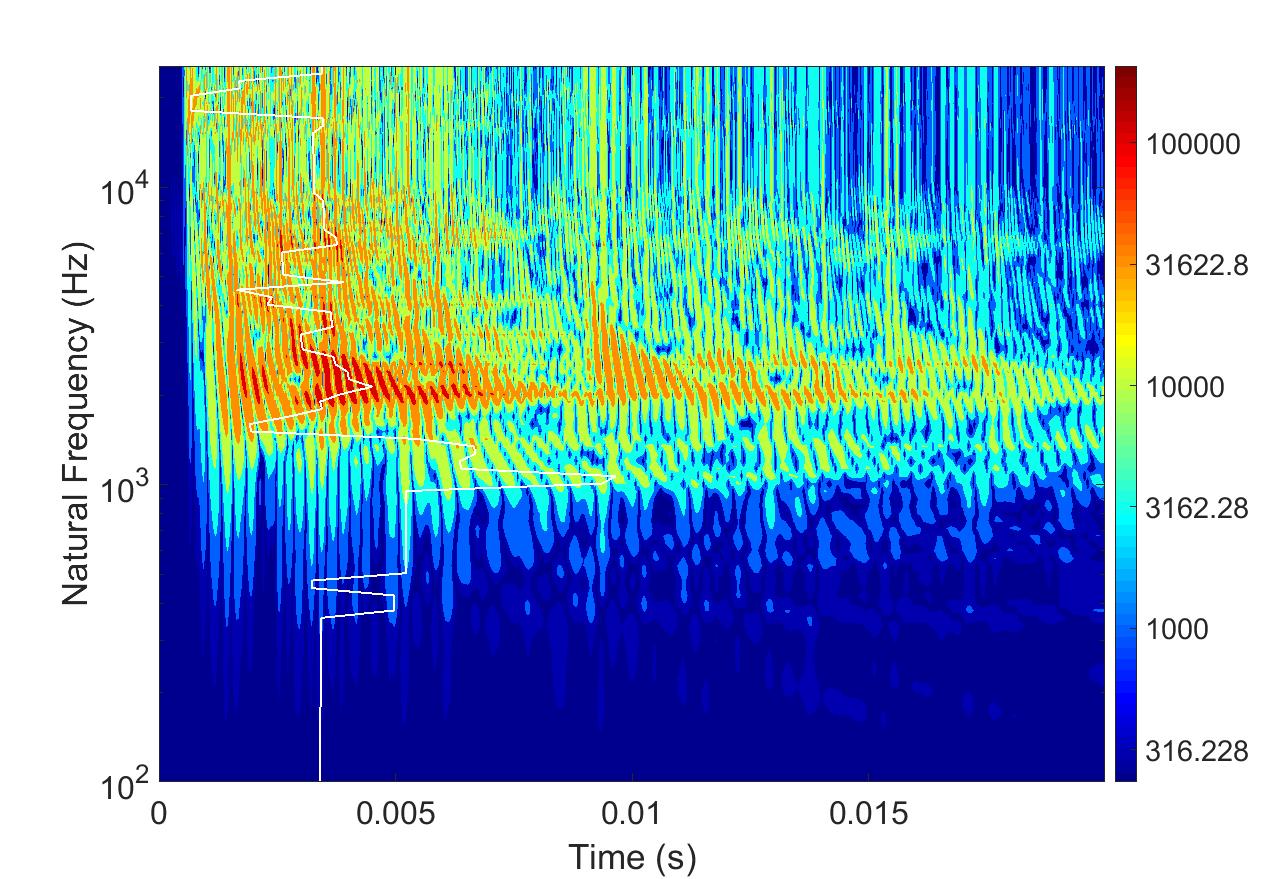}
		\caption{SRC of `MIS'}
		\label{mis_src}
	\end{subfigure}
	\begin{subfigure}{0.48\linewidth}
		\includegraphics[width=\linewidth]{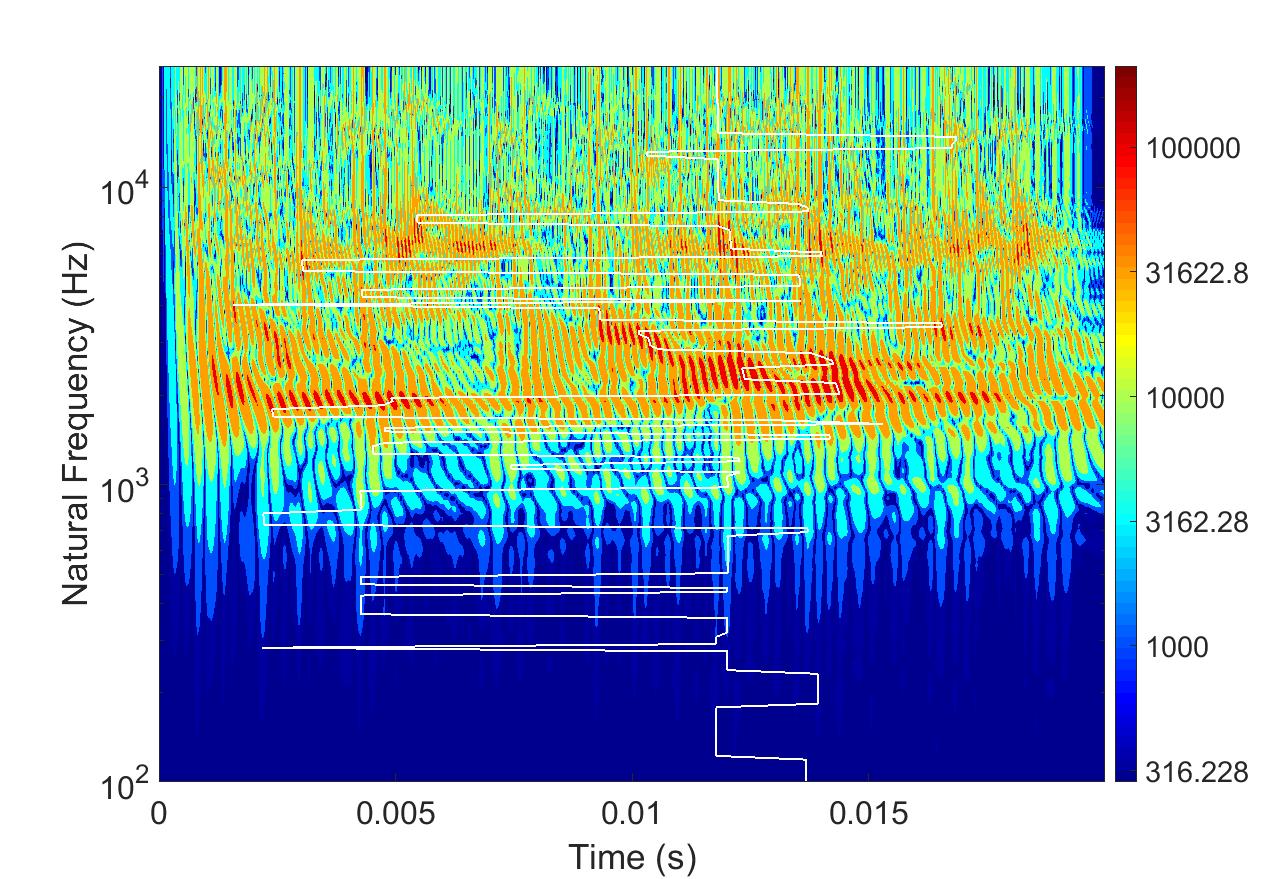}
		\caption{SRC of `RAS'}
		\label{ras_src}
	\end{subfigure}
	\begin{subfigure}{0.48\linewidth}
		\includegraphics[width=\linewidth]{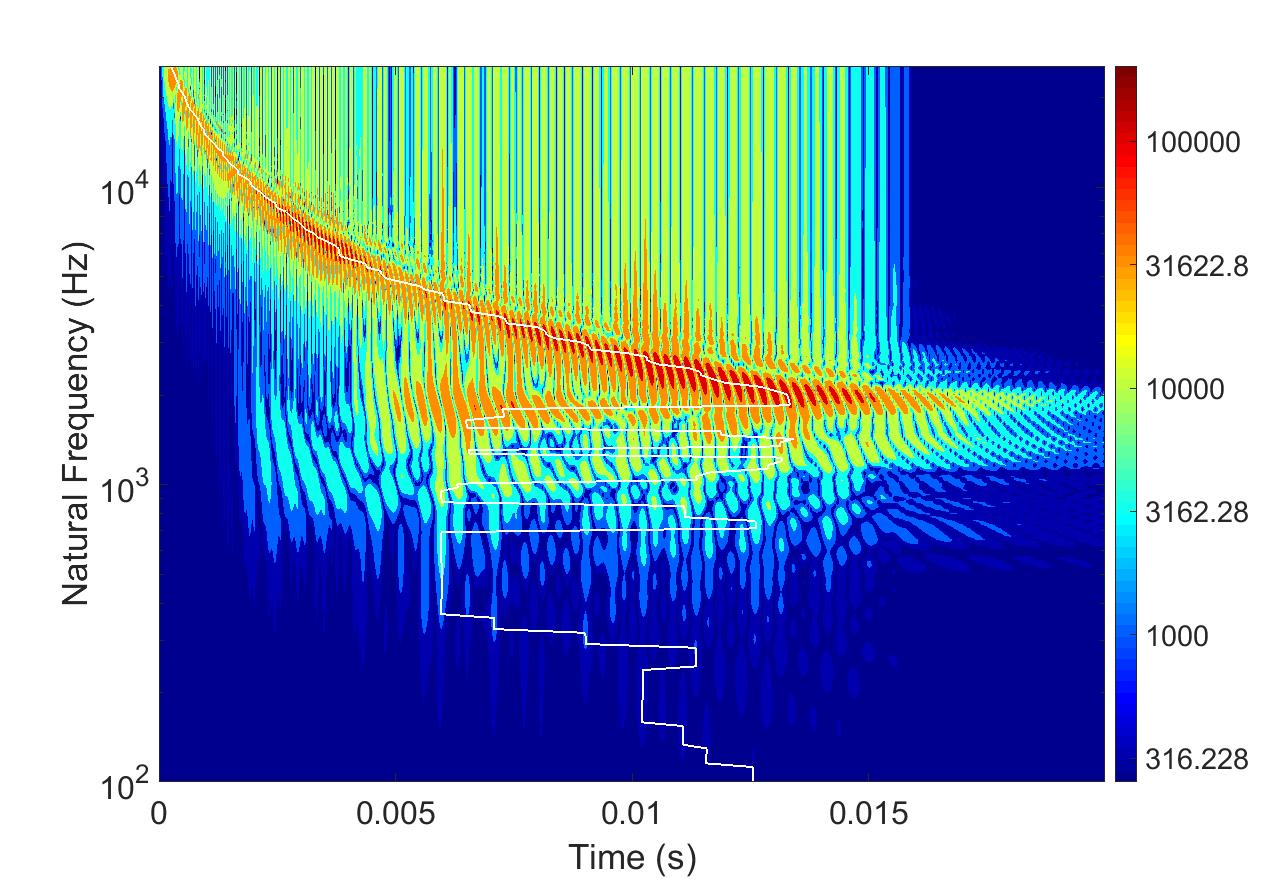}
		\caption{SRC of `RSS'}
		\label{rss_src}
	\end{subfigure}
	\caption{Comparison of SRC plots of various shock signals}
	\label{src_cwt}
\end{figure}

Although these shock signals have almost the same SRS curves, they represent three very different types of shock.
Their time-frequency features are also very different, as shown by their SRC (Q=10) in Figs.\ref{rvs_src_cwt}, and \ref{src_cwt}, respectively.
Therefore, their effects on mechanical systems should be generally different, which, however, can not be indicated by their SRS curves.

\begin{figure}[t]
	\centering
	\begin{subfigure}[b]{0.49\linewidth}
		\includegraphics[width=\linewidth]{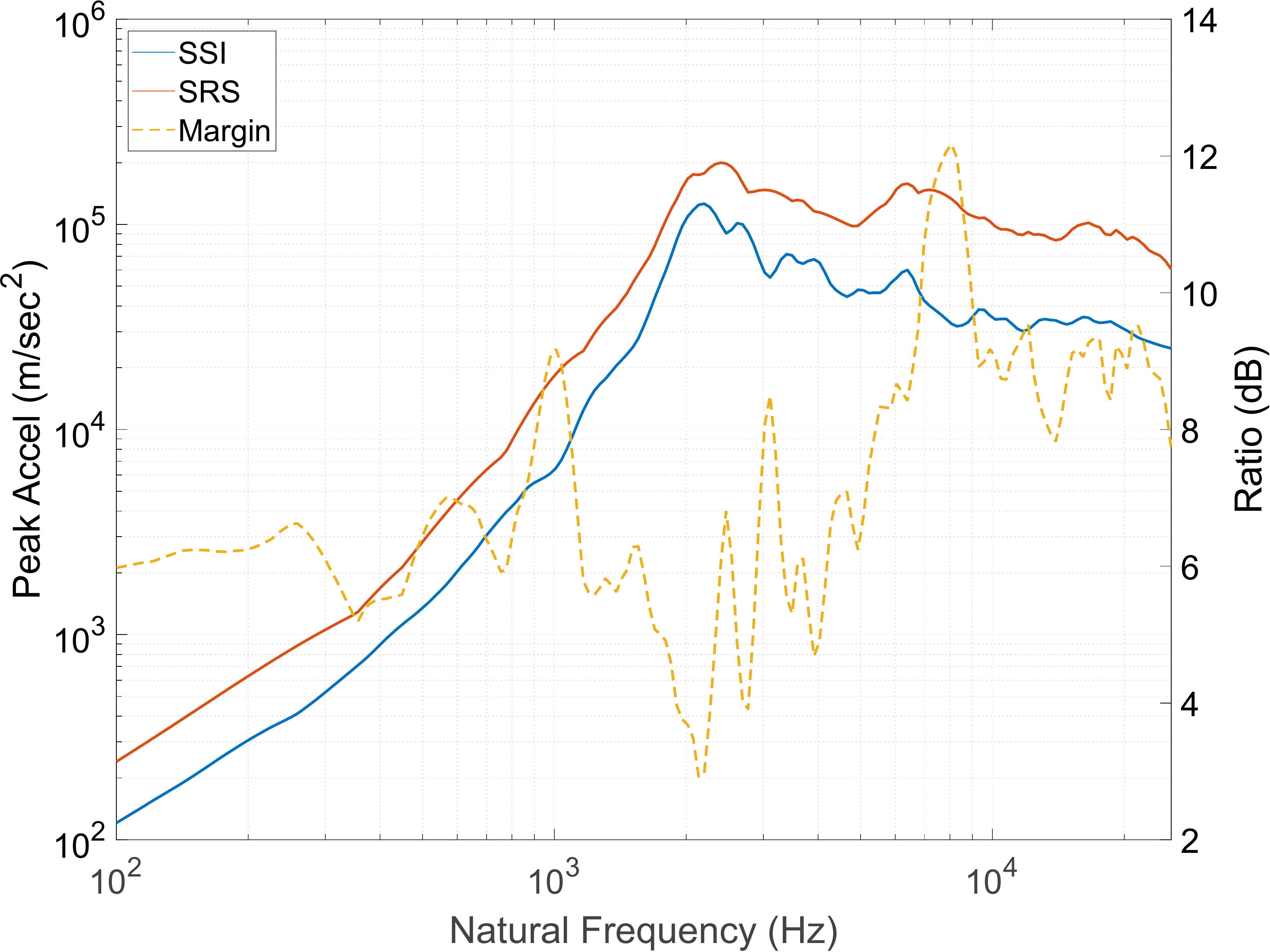}
		\caption{`RVS'}
	\end{subfigure}
	\hfill
	\begin{subfigure}[b]{0.49\linewidth}
		\includegraphics[width=\linewidth]{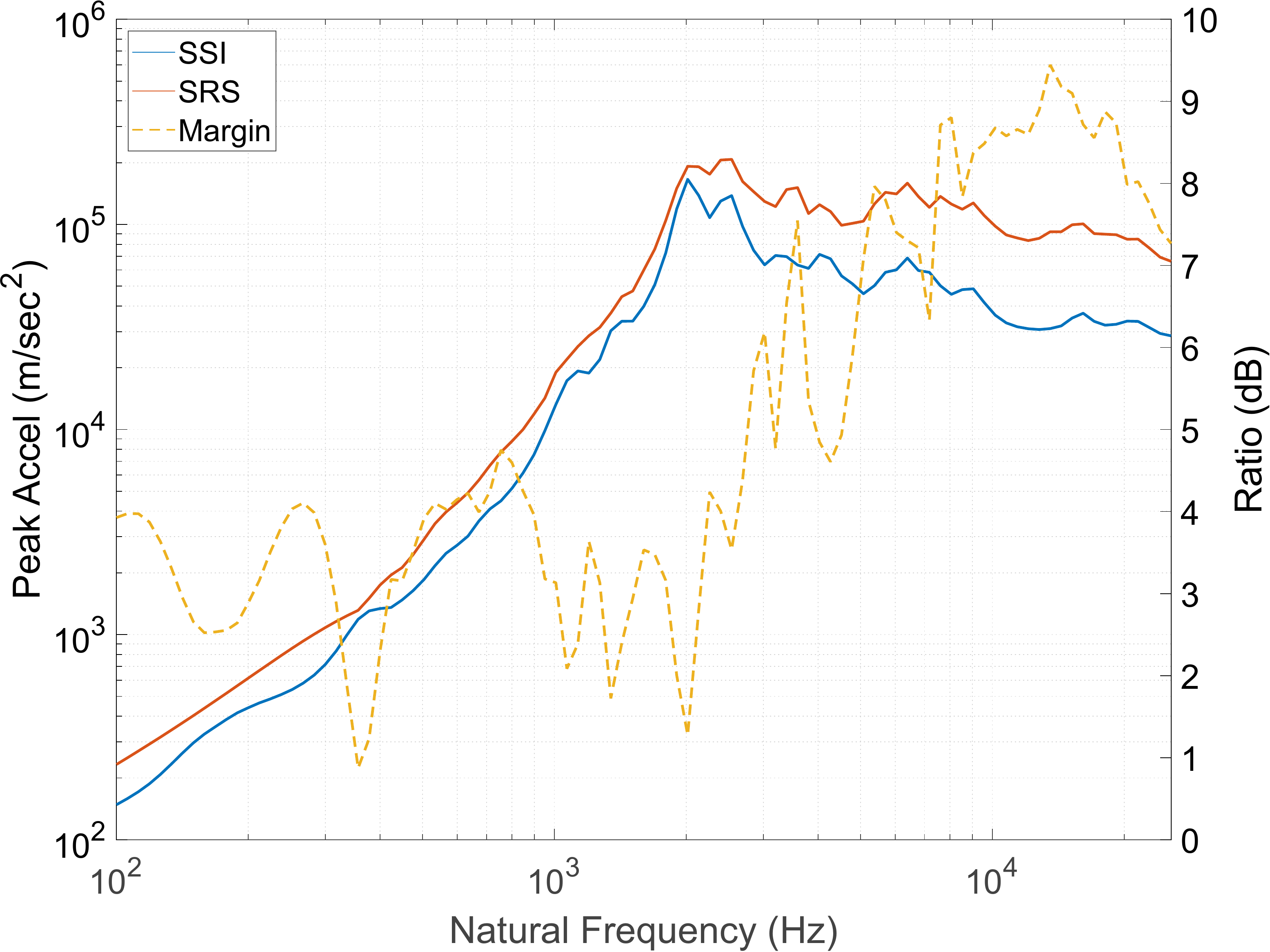}
		\caption{`MIS'}
	\end{subfigure}
	\begin{subfigure}[b]{0.49\linewidth}
		\includegraphics[width=\linewidth]{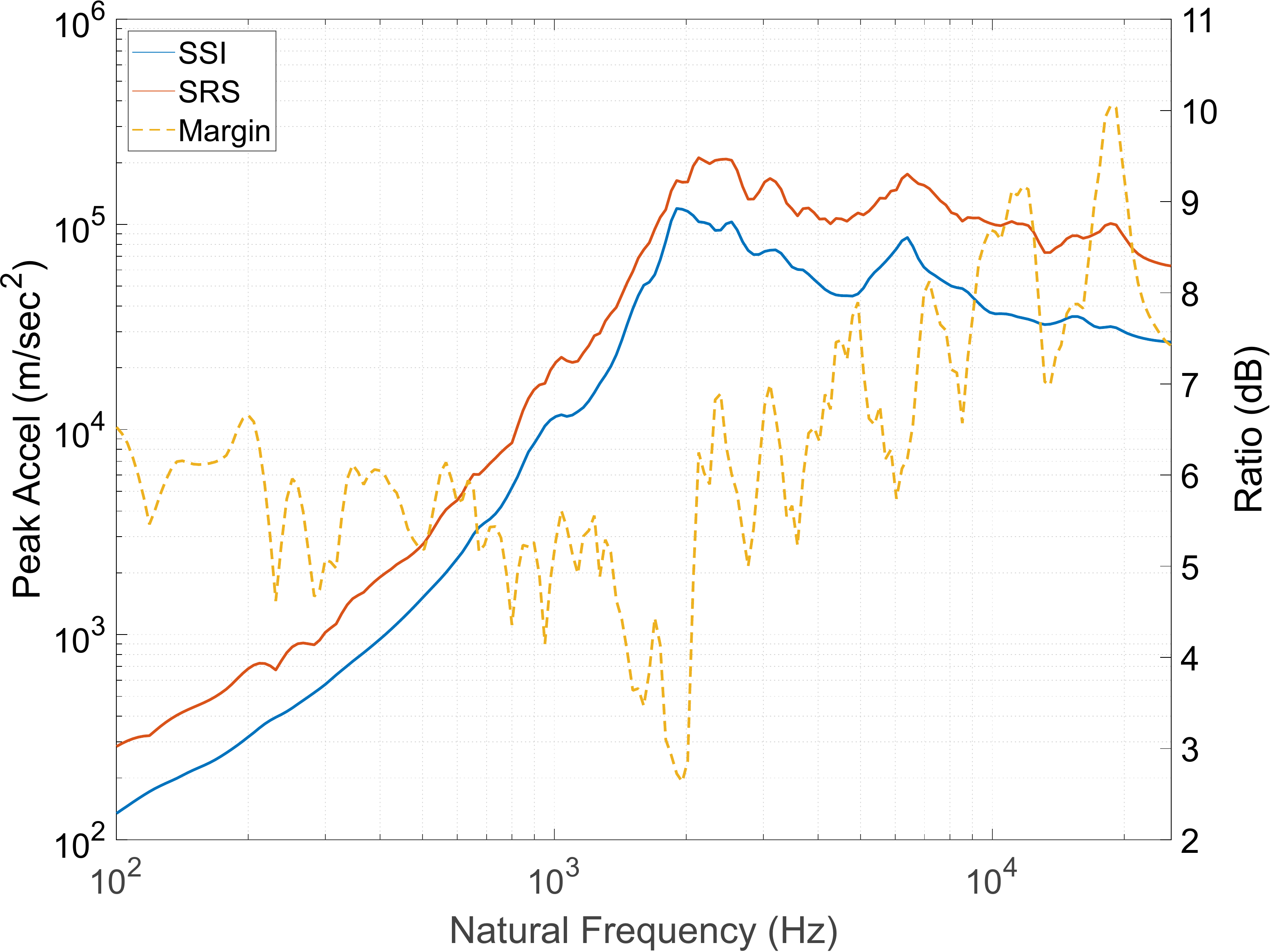}
		\caption{`RAS'}
	\end{subfigure}
	\hfill
	\begin{subfigure}[b]{0.49\linewidth}
		\includegraphics[width=\linewidth]{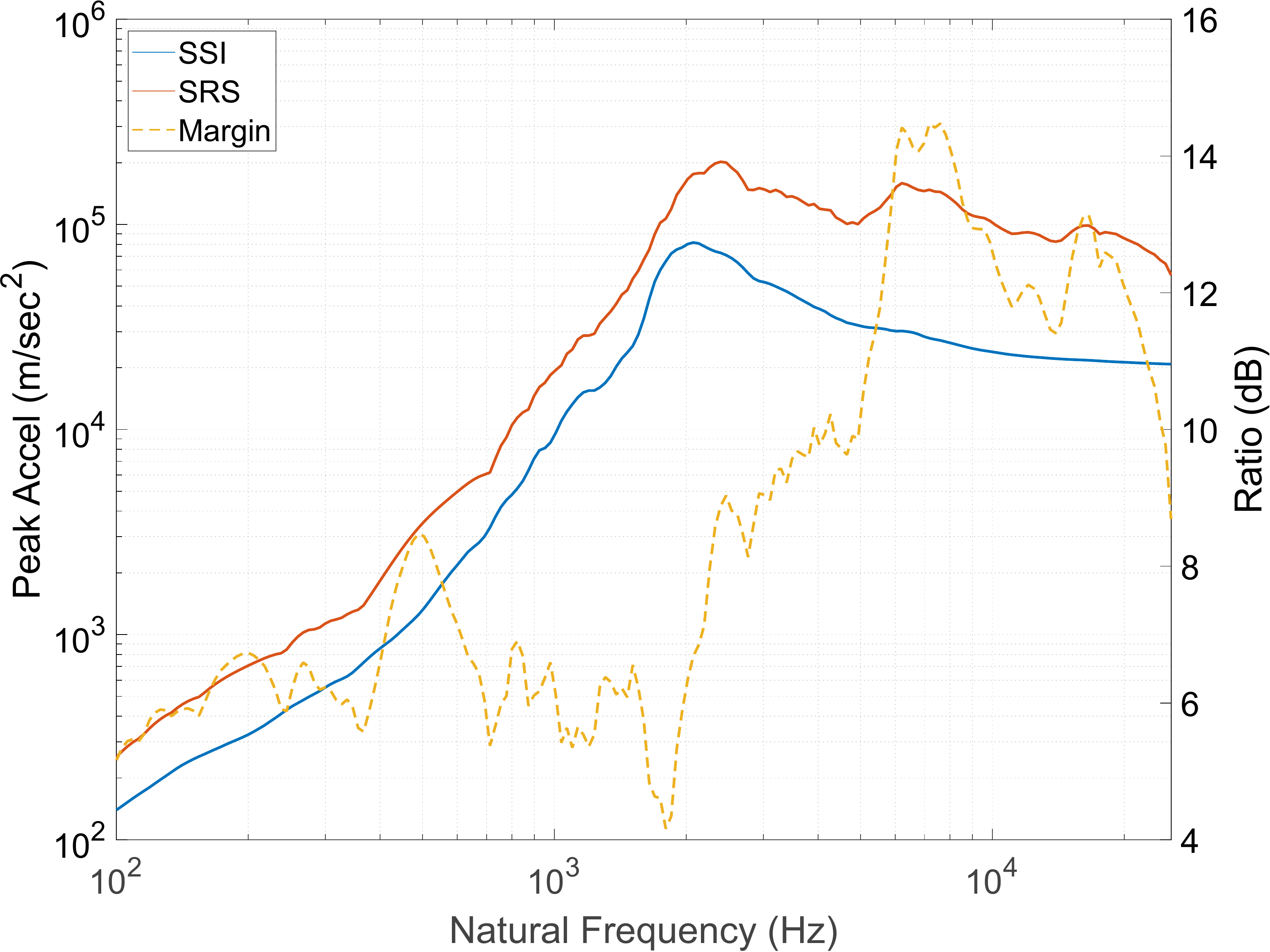}
		\caption{`RSS'}
	\end{subfigure}
	\caption{Dual spectra and margin $\bm{l}$ of four different shock signals}
	\label{dual_compare}
\end{figure}

The proposed dual spectrum method can improve the ordinary SRS method by providing the infimum of maximum shock responses for any given structures.
Fig.\ref{dual_compare} shows the dual spectrum and margin $\bm{l}$ of these four shocks.
The upper solid lines are ordinary SRS curves, while the lower solid lines are the proposed SSI spectrum extracted from their corresponding SRCs using Eq.(\ref{svd_decomposition}).
The curves of margin $\bm{l}$ are depicted by the dashed line with their units on the right axis.

Although the same supremum is indicated by their respective SRS curves, these four shocks have different margins and SSI curves.
For convenience, the entire frequency domain is divided into low-, mid- and high-frequency ranges by taking the knee frequency as a reference, which is about 2000 Hz in this case.
Generally, the margins in low- and mid-frequency ranges are lower than the margins in high-frequency ranges.
The $\bm{l}$ value can be used to estimate the necessary amplification to conduct fully conservative testing in a laboratory environment.
In low- and mid-frequency ranges, the margins of shocks `MIS' and `RAS' are about 3dB and 6dB, respectively, which can be linked to the 3dB/6dB margin in the 810G standard.
While in the high-frequency range, these margin increase exponentially (i.e. the trend is nearly linear in logarithm coordinate in Fig.\ref{dual_compare}) to 9 dB.
The margin of RSS is much higher than those of other shocks, especially in mid- and high-frequency ranges.
The large margin for `RSS' indicates a large uncertainty if `RSS' signal is used for shock severity test even its SRS is almost the same as the SRS of a field shock signal.

\subsection{FEM Validation}

\begin{figure}
	\centering
	\includegraphics[width=0.75\linewidth]{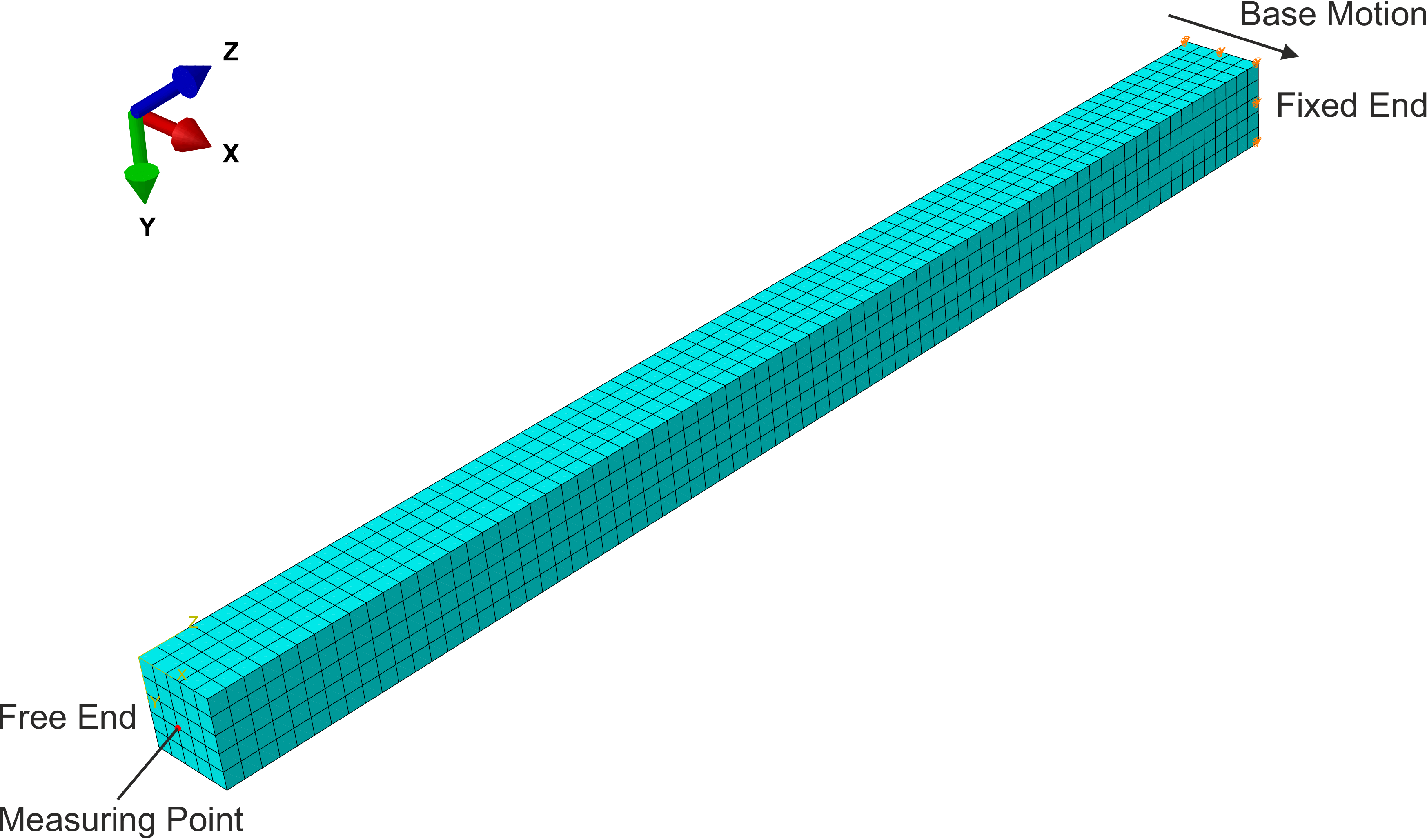}
	\caption{Illustration of FEA model}
	\label{cantilever_beam}
\end{figure}

In order to compare the shock severity using various methods, the response of a simple structure, i.e., a cantilever beam, is investigated using commercial FEA as an example.
The transient modal dynamic analysis solver in Abaqus 6.14-3 is used in this simple case.
As shown in Fig.\ref{cantilever_beam}, acceleration response is measured at the centre of the free end of the cantilever beam, which is excited by the base motion (shocks) applied at the fixed end in the transverse section direction (i.e. x-direction).
This $150\times10\times10$ mm beam is made of aluminium, whose density, Young's modulus, Poisson's ratio and dilatation wave speed are 2700 kg/m$^3$, 68.9 GPa, 0.32 and 6043 m/s, respectively.
Lanczo's eigensolver is adopted to obtain the natural frequencies up to 25,600 Hz, corresponding to the maximum natural frequency in the SRS shown in Fig.\ref{srs_compare}.
Time increment at $5\times 10^{-6}$ s and mesh size of 2 mm are adopted to meet the simulation requirements\cite{YAN2018}.
To be consistent with the Q=10 damping during SRS calculation, the damping ratio of 0.05 is used in the modal dynamic analysis.
Table \ref{modal_information} shows the modal information of the beam in the transverse direction of the beam.
The accumulated effective mass of all calculated modes is 37.98 g, which is already 93.78\% of the total mass of 40.50 g.

\begin{table}
	\centering
	\begin{threeparttable}
		\caption{Modal information for the cantilever beam}
		\label{modal_information}
		\small{
			\begin{tabular}{lllll}
				\hline
				Mode No. & $f$ (Hz) & $\Gamma$ & $\Phi$ & $M_{\text{eff}}$ (kg) \\
				\hline
				1  & 355 & 0.0965    & 6.084      & 0.0093  \\
				2  & 355 & 0.1244     & 7.841      & 0.0154   \\
				3  & 2183 & -0.0306   & 3.409      & 0.0009 \\
				4  & 2183 & -0.0822   & 9.170      & 0.0067  \\
				5  & 4531 & 4.104-12  & 1.656      & 1.684e-23 \\
				6  & 5932 & -0.0499   & -9.237     & 0.0024  \\
				7  & 5932 & 0.0136    & 2.510      & 0.0002 \\
				8  & 8437 & 3.470-14  & -7.268e-06 & 1.204e-27 \\
				9  & 11168 & 0.0092   & -2.294     & 8.465e-05 \\
				10 & 11168 & -0.0361   & 9.010      & 0.0013  \\
				11 & 13591 & -3.508e-12 & -1.656     & 1.230e-23 \\
				12 & 17621 & -0.0287   & -8.837     & 0.0008 \\
				13 & 17621 & 0.0055   & 1.677      & 2.975e-05 \\
				14 & 22640 & -5.409e-12 & 1.65      & 2.926e-23 \\
				15 & 25020 & -0.0219   & 7.88      & 0.0005 \\
				16 & 25020 & 0.0101    & -3.644     & 0.0001 \\
				17 & 25288 & -4.333e-13 & -7.279e-05 & 1.877e-25\\
				\hline
			\end{tabular}
		}
		\begin{tablenotes}
			\item Note: $\Gamma$ is the modal participation factor; $\Phi$ is the mass normalized mode shape; $M_{\text{eff}}$ is the effective mass.
		\end{tablenotes}
	\end{threeparttable}
\end{table}

\begin{table}[htpb!]
	\centering
	\caption{Maximum acceleration responses (m/s$^2$) of the cantilever beam from various methods}
	\label{response}
	\begin{tabular}{ c c c c c c}
		\hline 
		Shock\textbackslash Method & FEM & $\maxnorm{\bm{M} \bm{x}_i}$ & $\maxnorm{\bm{N}\bm{x}}$ & $\bm{v}_\SSI^\top \bm{x}$ & $\bm{v}_\SRS^\top \bm{x}$ \\ 
		\hline 
		RVS & 1.90$\times 10^5$ & 1.82$\times 10^5$ & 2.11$\times 10^5$ & 1.60$\times 10^5$ & 2.94$\times 10^5$ \\ 
		\hline 
		MIS & 1.78$\times 10^5$ & 1.78$\times 10^5$ & 2.00$\times 10^5$ & 1.63$\times 10^5$ & 2.94$\times 10^5$ \\
		\hline
		RAS & 2.24$\times 10^5$ & 2.15$\times 10^5$ & 2.29$\times 10^5$ & 1.51$\times 10^5$ & 3.25$\times 10^5$ \\
		\hline
		RSS	& 1.50$\times 10^5$ & 1.46$\times 10^5$ & 1.75$\times 10^5$ & 1.01$\times 10^5$ & 2.94$\times 10^5$ \\
		\hline
	\end{tabular} 
\end{table}

\begin{figure}
	\centering
	\includegraphics[width=0.9\linewidth]{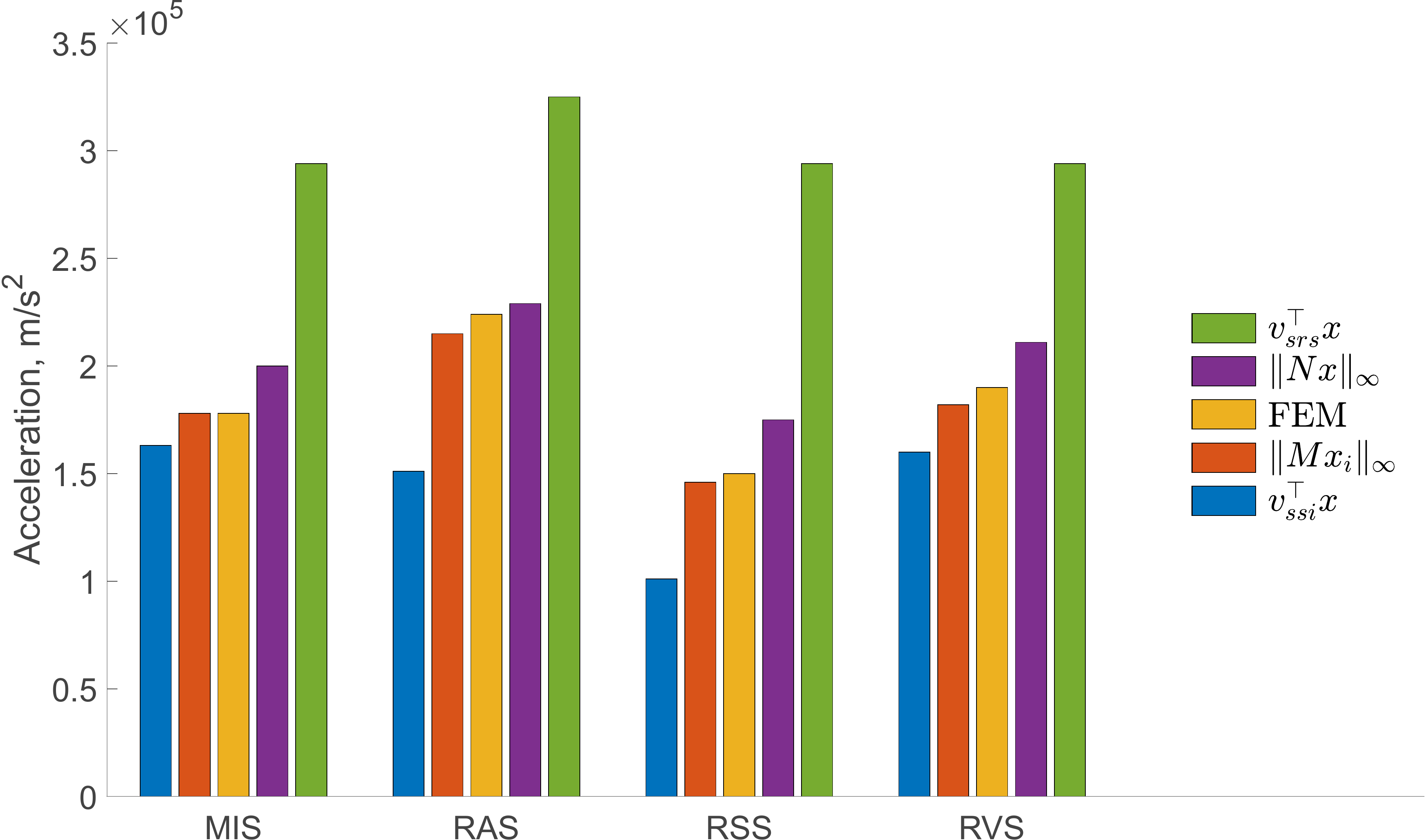}
	\caption{Magnitude relationship diagram of maximum acceleration responses of the cantilever beam subjected to various shocks with the same SRS}
	\label{bound}
\end{figure}

\begin{figure}
	\centering
	\includegraphics[width=0.7\linewidth]{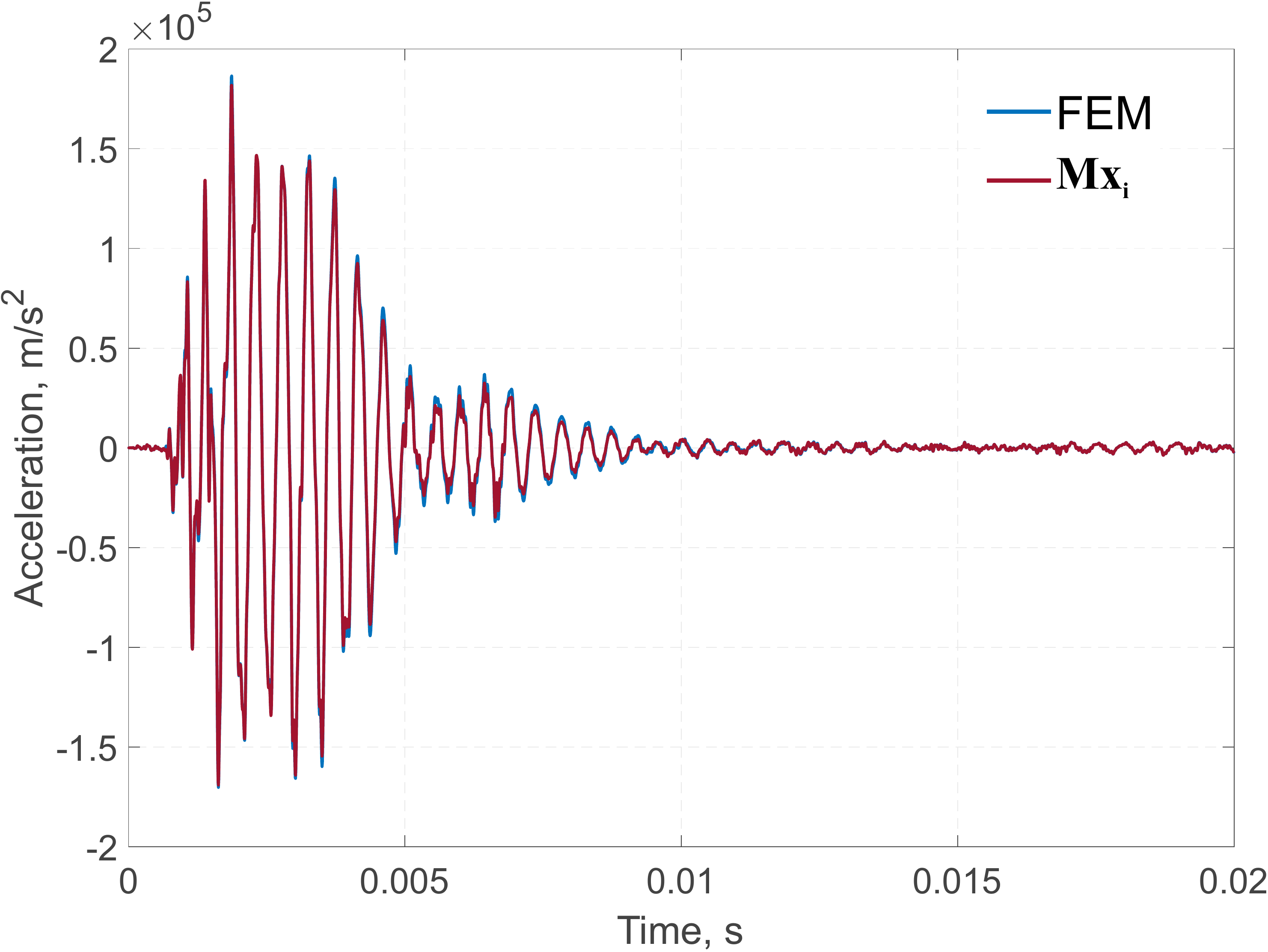}
	\caption{Acceleration response of the cantilever beam under RVS calculated by $\bm{M}\bm{x}_i$ and FEM}
	\label{fem_mxi}
\end{figure}

The absolute maximum acceleration response of the cantilever beam at the measuring point under different shocks are calculated from various methods based on the modal information in Table \ref{modal_information}.
The results are recorded in Table \ref{response}, whose magnitude relationship diagrams are illustrated in Fig.\ref{bound}.
The acceleration-time histories in Fig.\ref{fem_mxi} obtained from FEM and $\bm{M}\bm{x}_i$ are consistently close, as expected, because they both give the actual acceleration responses in the structure.
Small differences between their maximum responses calculated from $\maxnorm{\bm{M} \bm{x}_i}$ and FEM mainly come from the inaccuracy of the data extraction from the linear interpolation of matrix $\bm{M}$.
Large response differences are observed among different shocks, as shown in Table \ref{response}, although these shocks have almost the same SRS.
Due to the phase difference, the actual maximum response excited by `RAS' is 49.33\% (3.48 dB) higher than that for `RSS', which demonstrates the uncertainty of SRS method.
The maximum response provided by $\maxnorm{\bm{N}\bm{x}}$ is slightly higher than that given by $\maxnorm{\bm{M}\bm{x}_i}$, as shown in Eq.(\ref{inequality1}).
In the diagram of magnitude relationship, the responses calculated by SSI and SRS methods successfully bound the actual maximum response as the infimum and supremum, which is consistent with Eq.(\ref{inequality2}).
Furthermore, the actual maximum response is closer to the infimum response determined by SSI than to the ordinary supremum response determined by SRS, especially for the pyroshock and mechanical shocks.

\begin{figure}[!htbp]
	\centering
	\includegraphics[width=\linewidth]{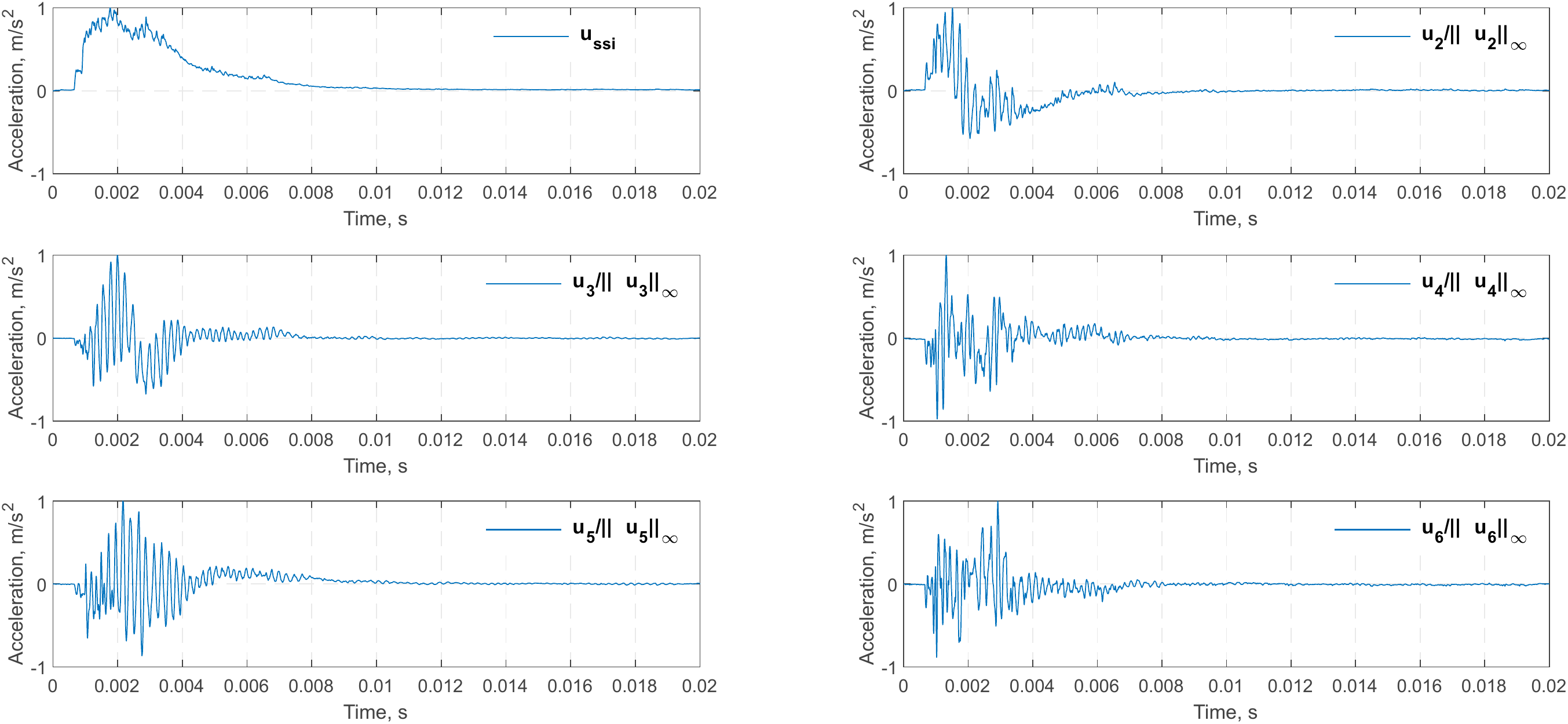}
	\caption{Weighted singular vectors of RVS in time domain}
	\label{time_subplot}
\end{figure}

\begin{figure}[htpb!]
	\centering
	\begin{subfigure}{0.45\linewidth}
		\includegraphics[width=\linewidth]{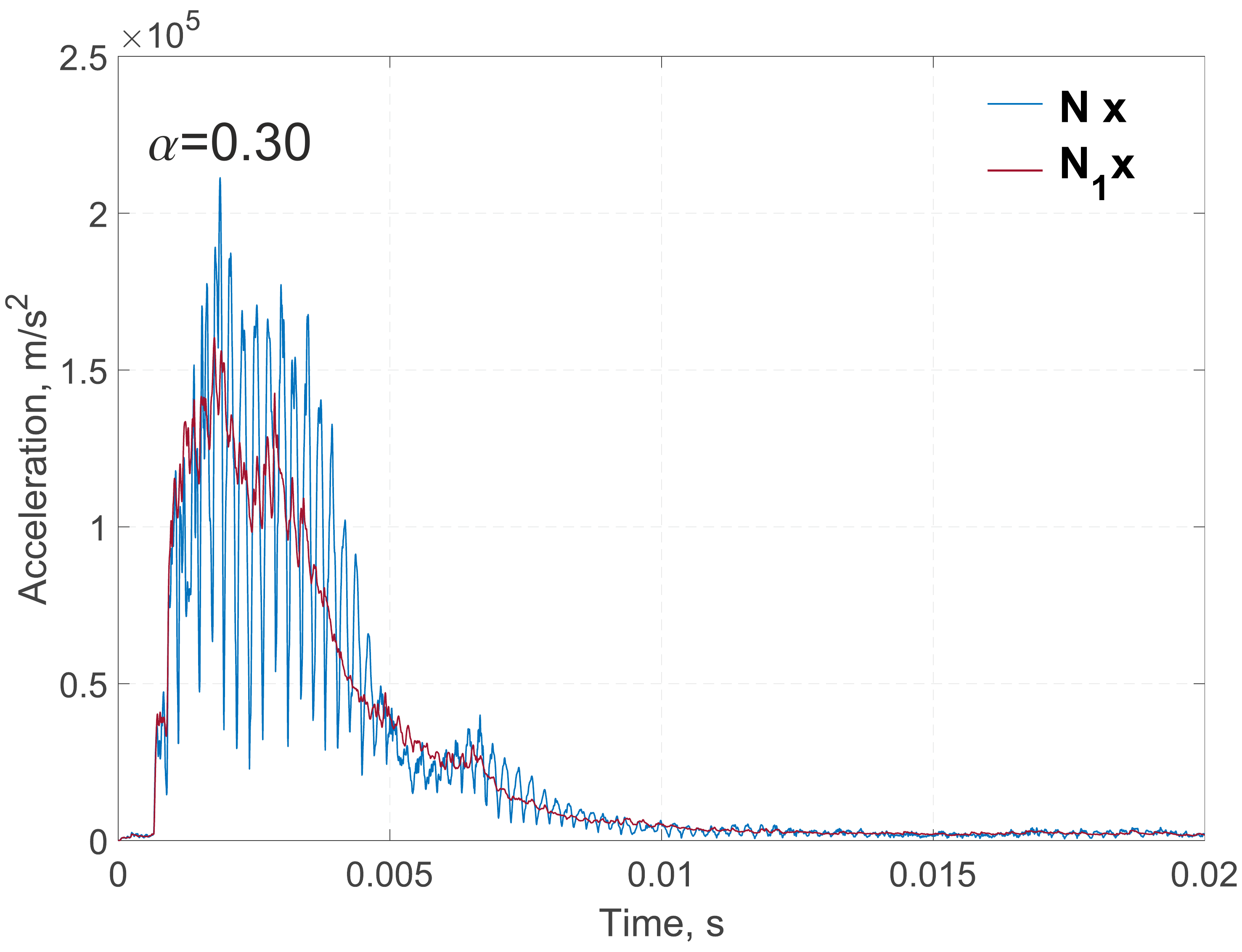}
		\caption{RVS}
	\end{subfigure}
	\hfill
	\begin{subfigure}{0.45\linewidth}
		\includegraphics[width=\linewidth]{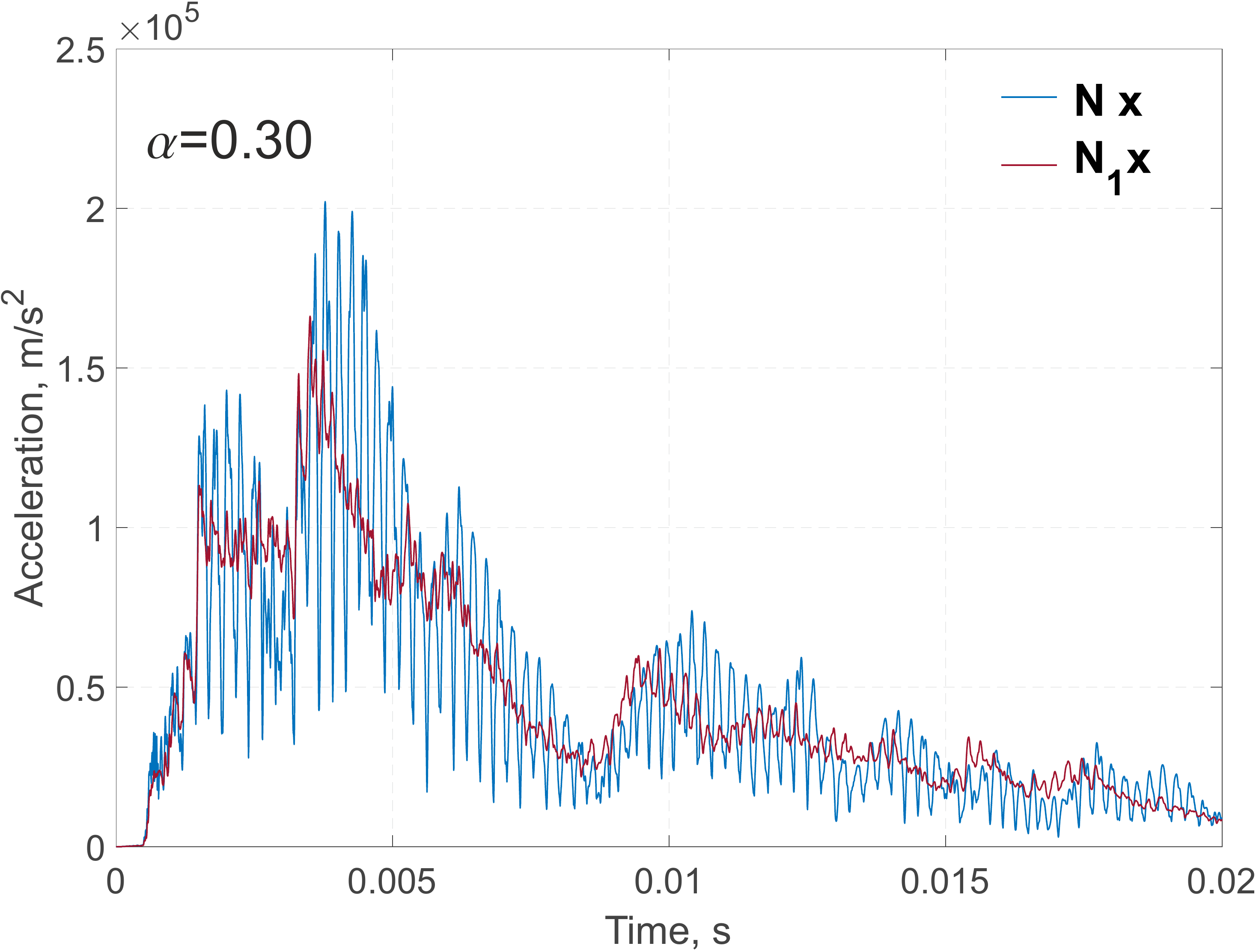}
		\caption{MIS}
	\end{subfigure}
	\begin{subfigure}{0.45\linewidth}
		\includegraphics[width=\linewidth]{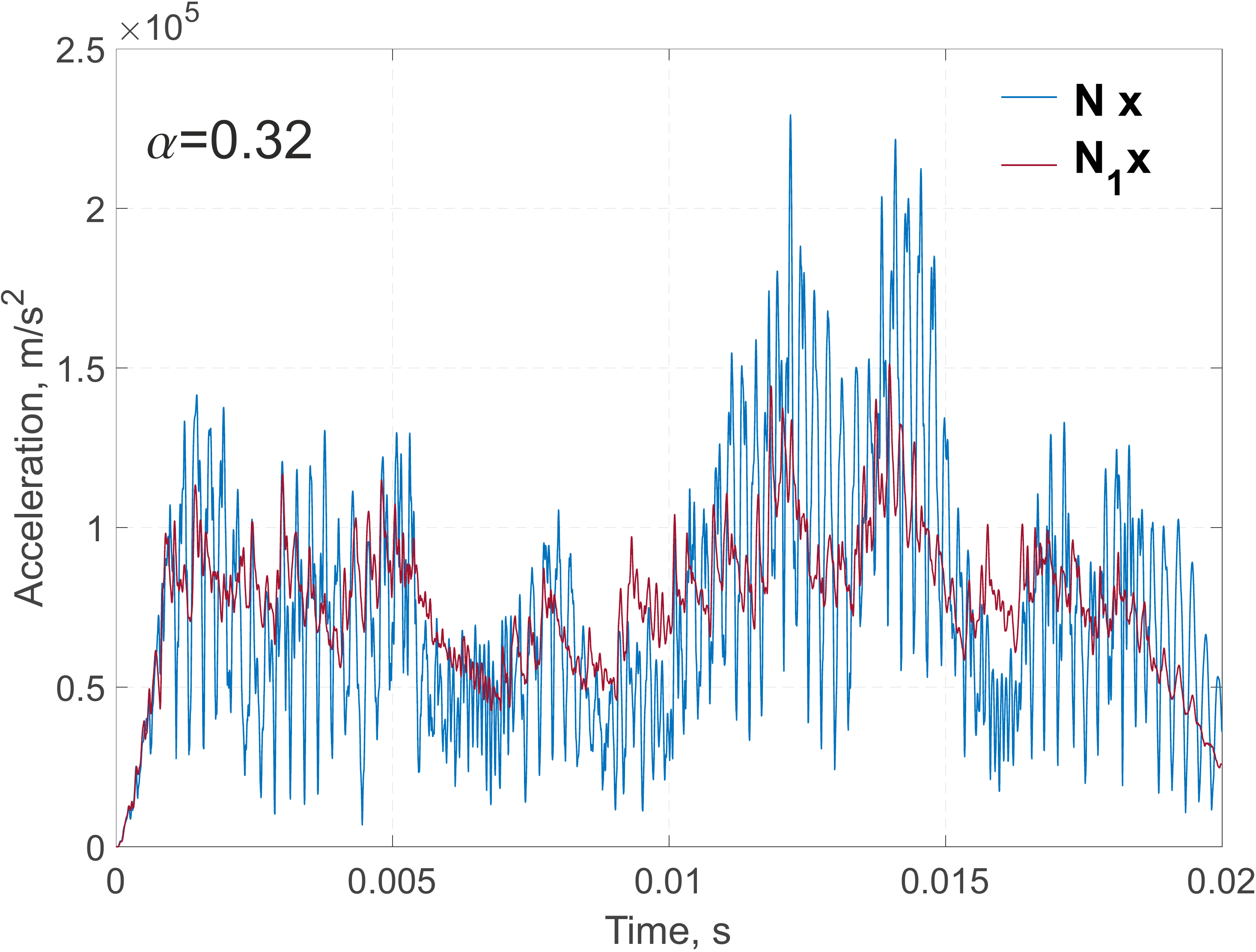}
		\caption{RAS}
	\end{subfigure}
	\hfill
	\begin{subfigure}{0.45\linewidth}
		\includegraphics[width=\linewidth]{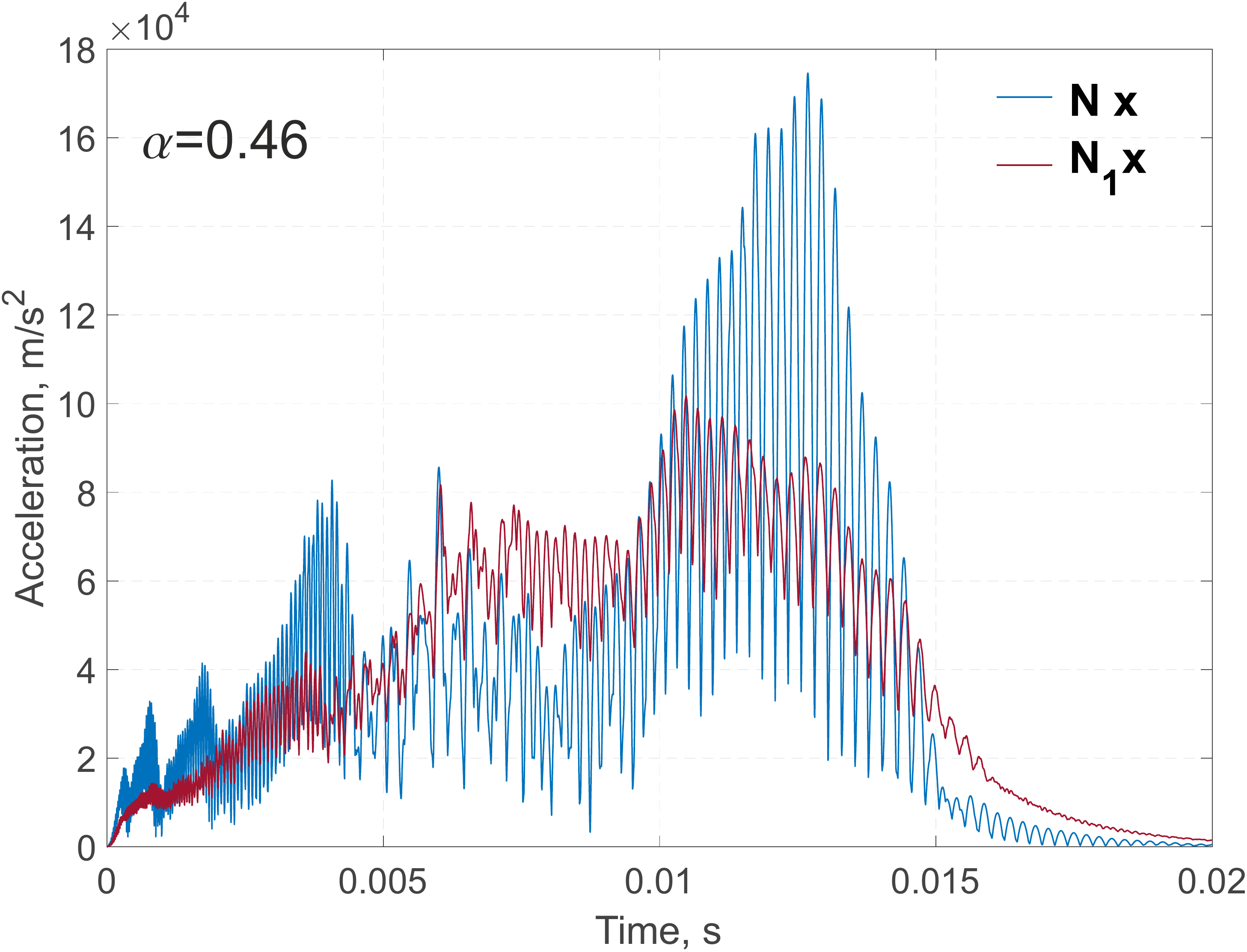}
		\caption{RSS}
	\end{subfigure}
	\caption{Comparison of shock responses calculated by $\bm{N}\bm{x}$ and $\bm{N}_1\bm{x}$}
	
	\label{nx_n1x}
\end{figure}

The first 6 singular vectors ($\bm{u}_k$) of RVS in time domain normalized by their corresponding maximum norm ($\maxnorm{\bm{u}_k}$) are listed in Fig.\ref{time_subplot}, where the first normalized singular vector is the dominant time history component $\bm{u}_{\SSI}$.
The comparison of shock responses of various shocks calculated by $\bm{N}\bm{x}$ and $\bm{N}_1\bm{x}$ are illustrated in Fig.\ref{nx_n1x} along with their corresponding relative residual error $\alpha$.
Since $\bm{N}$ is the non-negative matrix, the linear combination of $\bm{N}$'s columns, i.e. $\bm{N}\bm{x}$, is also a non-negative vector.
It is evident that $\bm{N}_1 \bm{x}$ approaches $\bm{N} \bm{x}$ in terms of signal energy without retaining higher frequency oscillations, which are included in the rest of higher order matrix components.
The residual error $\alpha$ can indicate signal energy difference and its representativeness of $\bm{N}_1 \bm{x}$.
For shock events RVS, MIS and RAS with small $\alpha$ ($\sim 30$), it is reasonable to consider SRC as separable functions, and the response by $\bm{N}_1 \bm{x}$ can provide average trends; while for event RSS with a large $\alpha$ (=0.46), only $\bm{N}_1 \bm{x}$ is insufficient to characterise the trend of $\bm{N} \bm{x}$.

\section{Conclusions}

This study provides a further theoretical basis for the measure of shock severity.
The traditional shock response spectrum (SRS), as the supremum of the shock response, is insufficient to describe the severity of a shock signal.
Based on the responses of single degree of freedom (SDOF) oscillators to a given shock signal, shock response matrix (SRM) is introduced to record a complete set of responding data of SDOF oscillators in both time and frequency domains.
It is found that the SRC has rich information and can be used as a time-frequency analytical tool in the analyses of mechanical shock signals.
By applying singular value decomposition (SVD) theorem to the SRM, the first singular vector is extracted as shock severity infimum (SSI) based on the conventional SRS concept and calculation process.
The range of the severity of any given shock environment can be explicitly bounded by the dual spectra defined by SRS and SSI curves, which is illustrated by the FEA simulations of a representative structure.
To support the shock resistance design of devices and equipment, the proposed dual spectra can also be used to design more reliable shock tests in laboratory environment since the reproduced shocks in laboratory environments cannot always represent the field shock signals even they have similar SRS curves.
The dual spectra allow the definition of the SRS margin for a given shock specified in terms of its SRS, within which the simplified shock excitations in laboratory environments can be used to substitute field shock signals to certify the devices and equipment. 

\enlargethispage{20pt}


\dataccess{The time histories of four shock signals, the source code of SSI method and the modal information of the cantilever beam are released on the GitHub of the author: \href{https://github.com/galois-yan/Shock_Severity_Infimum_Method}{https://github.com/galois-yan/Shock\_Severity\_Infimum\_Method}.}

\aucontribute{Y.Y. and Q.L. designed research; Y.Y. performed research; Y.Y. contributed new analytic tools; Y.Y. analysed data; Q.L. guided the research; Y.Y. and Q.L. wrote the paper.
Both authors agreement to be accountable for all aspects of the work.}

\competing{The authors declare no competing interests.}

\funding{The authors received no funding for this work.}



\appendix

\section{Proof of Eq.(\ref{trend})}\label{proof_trend}

The operator norm of a matrix $\bm{A}$ is defined as:
\[ \| \bm{A} \|_{op} := \inf \{ c\geq0 : \|\bm{A}\bm{x}\|_2 \leq c\|\bm{x}\|_2 \quad \text{for all }\bm{x}\in \bm{R}^n \} \]
Re-organise Eq.(\ref{trend}), we have:
\[ \| \bm{N}\bm{x}-(\sigma_1 \bm{v}_1 ^\top \bm{x}) \bm{u}_1 \|_2 =\|\bm{N} \bm{x}-\bm{N}_1 \bm{x}\|_2 =\|(\bm{N}-\bm{N}_1) \bm{x}\|_2  \]
The Eckart-Young-Mirsky theorem says that
\[ \|\bm{A}-\sum_{i=1}^{k}\bm{A}_i\|_{op}=\sigma_{i+1} \]
In the case of matrix $\bm{N}$ and $k=1$, we have
\[\|\bm{N}-\bm{N}_1 \|_{op} =\sigma_2 \]
According to the definition of operator norm, the inequality exist for arbitrary $\bm{x}$ that
\begin{equation*}
\begin{split}
& \|(\bm{N}-\bm{N}_1) \bm{x}\|_2 \leq \sigma_2\|\bm{x}\|_2\\
\Rightarrow \quad & \| \bm{N}\bm{x}-(\sigma_1 \bm{v}_1 ^\top \bm{x}) \bm{u}_1 \|_2 \leq \sigma_2 \| \bm{x} \|_2
\end{split}
\end{equation*}

\section{Proof of Eq.({\ref{inequality2}})}\label{proof_inequality2}

\begin{flalign*}
\text{(i)}&&\bm{v}_\SSI^\top\bm{x}&=\maxnorm{\bm{N}_1 \bm{x}}&
\end{flalign*}
\paragraph{Proof}
\[ \maxnorm{\bm{N}_1 \bm{x}} = \maxnorm{\sigma_1 \bm{u}_1 \bm{v}_1^\top \bm{x}}\]
The singular value $\sigma_1$ and the inner product $\bm{v}_1^\top \bm{x}$ are scalars, which can be collected outside the maximum norm operation,
\[\maxnorm{\sigma_1 \bm{u}_1 \bm{v}_1^\top \bm{x}}=\sigma_1 \maxnorm{\bm{u}_1} \bm{v}_1^\top \bm{x} =\bm{v}_\SSI^\top\bm{x}\]
according to Eq.(\ref{n1}).

\begin{flalign*}
\text{(ii)}&&\bm{v}_\SRS^\top\bm{x}  &\geq \maxnorm{\bm{N} \bm{x}}&
\end{flalign*}

\paragraph{Proof}
A SRS matrix $\bm{N}_{\SRS}$ can be constructed by
\[\bm{N}_{\SRS}=\bm{J}_{m,1}\bm{v}_{\SRS}^\top\]
where $\bm{J}_{m,1}$ is a $m$ rows all-ones vector
\[\bm{J}_{m,1}=(1,1, \cdots , 1)^\top.\]
According to the definition of SRS in Eq.(\ref{srs_definition}), matrix $\bm{N}_{\SRS}$ is entrywise equal or greater than matrix $\bm{N}$.
\[ \bm{N}_{\SRS}-\bm{N} \geq 0 \]
Since vector $\bm{x}$ is a non-negative vector, we have
\[ (\bm{N}_{\SRS}-\bm{N})\bm{x}  \geq 0\] 
\[ \bm{N}_{\SRS}\bm{x}\geq \bm{N}\bm{x} \]
Therefore,
\[ \maxnorm{\bm{N}_{\SRS}\bm{x}}=\bm{v}_{\SRS}^\top \bm{x} \geq \maxnorm{\bm{N}\bm{x}}. \]


\vskip2pc


%
%
%
%
%
%
%
%
%
\bibliographystyle{RS}
\bibliography{mybibfile}
\end{document}